\documentclass[12pt]{article}

\begin{document}
\begin{center}

{\bf{\Large Rotating Black Hole Solutions with Axion Dilaton and
Two Vector Fields and  Solutions with Metric and Fields of the
Same Form}}

\vspace{1cm}
E. Kyriakopoulos\footnote{E-mail: kyriakop@central.ntua.gr}\\
Department of Physics\\
National Technical University\\
157 80 Zografou, Athens, GREECE
\end{center}

\begin {abstract}

We present two rotating black hole solutions with axion $\xi$,
dilaton $\phi$ and two $U(1)$ vector fields. Starting from a
non-rotating metric with three arbitrary parameters, which we have
found previously, and applying the "Newman-Janis complex
coordinate trick" we get a rotating metric $g_{\mu\nu}$ with four
arbitrary parameters namely the mass $M$, the rotation parameter
$a$ and the charges electric $Q_E$ and magnetic $Q_M$.  Then we
find a solution of the equations of motion having this
$g_{\mu\nu}$ as metric.  Our solution is asymptotically flat and
has angular momentum $J=M a$, gyromagnetic ratio $g=2$, two
horizons, the singularities of the solution of Kerr, axion and
dilaton singular only when $r=a\cos\theta=0$ etc.  By applying to
our solution the $S$-duality transformation we get a new solution,
whose axion, dilaton and vector fields have one more parameter.
The metrics, the vector fields and the quantity
$\lambda=\xi+ie^{-2\phi}$ of our solutions and the solution of :
Sen for $Q_E$, Sen for $Q_E$ and $Q_M$, Kerr-Newman for $Q_E$ and
$Q_M$, Kerr, Ref. \cite{Ky1}, Shapere, Trivedi and Wilczek (STW),
Gibbons-Maeda-Garfinkle-Horowitz-Strominger (GM-GHS),
Reissner-Nordstr\"{o}m,Schwarzschild are the same function of $a$,
and two functions $\rho^2=r(r+b)+a^2\cos^2\theta$ and
$\Delta=\rho^2-2Mr+c$, of $a$, $b$ and two functions for each
vector field , and of $a$, $b$ and $d$ respectively, where $a$,
$b$, $c$ and $d$ are constants. From our solutions several known
solutions can be obtained for certain values of their parameters.
It is shown that our two solutions satisfy the weak the dominant
and the strong energy conditions outside and on the outer horizon
and that all solutions with a metric of our form, whose parameters
satisfy some relations satisfy also these energy conditions
outside and on the outer horizon. This happens to all solutions
given in the Appendix. Mass formulae for our solutions and for all
solutions
 which are mentioned in the paper are given. One
mass formula for each solution is of Smarr's type and another a
differential mass formula. Many solutions with metric, vector
fields and $\lambda$ of the same functional form, which include
most physically interested and well known solutions, are listed in
an Appendix.
\\
PACS number(s): 04.20.Jb, 04.70.Bw

\end {abstract}

\section{Introduction}

It is well known that the solution of Kerr \cite{Ke1} can be
"derived" from the solution of Schwarzshild by means of the
"Newman and Janis complex coordinate trick" \cite{Ne1}. Later the
Kerr-Newman metric with electric charge only was derived from the
Reissner-Nordstr\"{o}m metric by the same trick and subsequently a
vector field was found, which together with the metric gives the
Kerr-Newnan solution of the Einstein-Maxwell equations \cite{Ne2}.
Recently it was shown that the metric of the axion-dilaton black
hole solution of Sen with electric charge only \cite{Se1} can be
derived from the metric of the
Gibbons-Maeda-Garfinkle-Horowitz-Strominger (GM-GHS)solution
\cite{Gi1} by the Newman and Janis method \cite{Ya1}. The GM-GHS
metric depends on two arbitrary parameters, while the metric of
Sen on three such parameters. The solution of Sen satisfies the
field equations of the Einstein-Maxwell-axion-dilaton gravity in
four dimensions, which are obtained from the action
\begin{equation}
S = \int d^{4}x \sqrt{-g}
\{R-2\partial_{\mu}\phi\partial^{\mu}\phi
-\frac{1}{2}e^{4\phi}\partial_{\mu}\xi\partial^{\mu}\xi-
e^{-2\phi}F_{\mu\nu}F^{\mu\nu} -\xi F_{\mu\nu}F^{\ast}_{\mu\nu}\}
\label{1-1}
\end{equation}
where $R$ is the Ricci scalar, $\xi$, $\phi$ and  $F_{\mu\nu}$ are
the axion the dilaton and a  $U(1)$ vector field respectively.

Non-rotating black hole solutions were found recently, whose
metric (for $c=0$, where $c$ is a parameter appearing in the
solutions) depends on three arbitrary parameters
\cite{Ky1}-\cite{Ky3}. One may ask if the rotating metric with
four parameters we get if we apply to this metric the Newman-Janis
trick can be the metric of a rotating black hole solution of the
equations of motion coming from the action (\ref{1-1}).  We tried
to do that but we failed. Subsequently we proved that this can be
done if we add to the Lagrangian of the action (\ref{1-1}) a
second $U(1)$ vector field. Actions of the above type with more
than one vector fields are frequently used, and it is argued that
"the presence of only one vector field is insufficient to generate
all interesting metrics" \cite{Be1}.

In Section 2 we shall apply to our metric the Newman-Janis method
and we shall derive a rotating black hole metric with four
parameters, which depends on the rotation parameter $a$ and two
functions $\Delta$ and $\rho^2$. This metric and the metrics of
all solutions, which are mentioned in this paper are the same
function of $a$, $\Delta$ and $\rho^2$.

In Section 3 starting from an action of the type (\ref{1-1}) but
with two vector fields $F_{\mu\nu}$ and $F'_{\mu\nu}$ we obtain
the equations of motion. Also we take
$F_{r\theta}=F_{t\phi}=F'_{r\theta}=F'_{t\phi}=0$ and we express
the rest four non-vanishing components of the field $F_{\mu\nu}$
in terms of two functions $\zeta$ and $\eta$ and the four
non-vanishing components of the field $F'_{\mu\nu}$ in terms of
two functions $\zeta'$ and $\eta'$ by the same ansatz. The vector
fields of all solutions  which are mentioned in this paper can be
expressed in terms of two functions by the same ansatz.

In Section 4 we calculate the components of the Ricci tensor
$R_{\mu\nu}$ using the metric we have found and we show that the
metric equations can be reduced to three relatively simple
equations.

In Section 5 we show that the equations of motion of the vector
fields can be solved exactly and we express the functions $\zeta$
and $\eta$ of the first vector field in terms of two arbitrary
functions of their arguments $G(r)$ and $H(y)$ and the functions
$\zeta'$ and $\eta'$ of the second vector field again in terms of
two arbitrary functions of their arguments $G'(r)$ and $H'(y)$.

The solution of the complex axion-dilaton equation of motion is
the most difficult part of the work. This is done in Section 6, in
which $\xi$, $e^{-2\phi}$, $\zeta$, $\eta$, $\zeta'$ and $\eta'$
are determined.

In Section 7 some physical characteristics of the solution are
determined. The mass, the angular momentum , the vector
potentials, the magnetic dipole moment and the gyromagnetic ratio
of the solution are calculated. The two infinite red shift
surfaces are obtained and it is shown that the solution has two
event horizons and a ring singularity, as the solution of Kerr
\cite{Ke1}, \cite{Ra1}. Also the area of the outer horizon, the
surface gravity, which is proportional to the Hawking temperature,
and the angular velocity are determined.

In Section 8 it is shown that for certain values of the four
arbitrary parameters of our solution a number of known solutions
or metrics of known solutions are obtained as special cases. These
are the solution of Sen for electric charge only \cite{Se1}, the
metric of the Kerr-Newman solution for equal electric and magnetic
charges \cite{Ca1}, the solution of Kerr \cite{Ke1}, a metric  we
have found previously \cite{Ky1}, \cite{Ky3}, the GM-GHS solution
\cite{Gi1}, the metric of the Reissner-Nordstr\"{o}m solution for
equal electric and magnetic charges and the Schwarzschild
solution.

In Section 9 by applying to the solution we have found the
S-duality transformation we obtain a ``new solution", which has
the same metric $g_{\mu\nu}$ but its axion, dilaton and vector
fields contain one more arbitrary parameter. Also it is pointed
out that the quantity $\lambda=\xi+ie^{-2\phi}$ of both solutions
we have found  takes the form of Eq (\ref{9-33}) and it is
expressed in terms of two real parameters $a$ ( it is
$y=a\cos\theta$ ) and $b$ and one real or complex parameter $d$. A
relation of the same form holds for all solutions with $\phi\neq
0$, which are mentioned in this paper.

In Section 10 we show that the solution of Sen with both charges
electric and magnetic \cite{Se2} and the solution of Shapere,
Trivedi and Wilczek (STW) \cite{Sh1} are special cases of the
``new solution".

In Section 11 we show that our two solutions and also all
solutions, whose metric is of the form considered in this paper
and its parameters satisfy some relations, satisfy the weak the
dominant and the strong energy conditions outside and on the outer
horizon. All solutions given in the Appendix of the paper satisfy
all energy conditions outside and on the outer horizon.

In section 12 we calculate a mass formula for the solutions with
metric of the form of  Eqs (\ref{2-18}), (\ref{2-24}) and
(\ref{2-25}) and angular momentum $J=Ma$, where $a$ is a non-zero
or zero constant. The formula is homogeneous in its arguments.
Applying to this expression Euler's theorem on homogeneous
functions we find a mass formula of Smarr's type \cite{Sm1}. Also
taking its differential we find a differential mass formula. The
mass formulae of our two solutions and all other solutions of the
Appendix with metric of this type are given. Also for solutions
with metric of the form of Eqs (\ref{2-18}) and (\ref{2-25'}) and
angular momentum $J=Ma$ a mass formula is obtained, which in the
case of the Kerr-Newman solution is homogeneous in its arguments.
From this expression mass formulae for the Kerr-Newman solution
with arbitrary $Q_E$ and $Q_M$ are calculated, whose special cases
are the mass formulae of the solution of Kerr and the mass
formulae of the solution of Reissner-Nordstr\"{o}m.

In the Appendix  we make a list of physically interesting
solutions, whose metric and fields are of the general form
considered in the paper. More specifically for all solutions of
the Appendix the metric $(g_{\mu\nu})$ is the same function of the
rotation parameter $a$ and two functions
$\rho^2=r(r+b)+a^2\cos^2\theta$ and $\Delta=\rho^2-2Mr+c$ with
$M$, $b$ and $c$ constants, the vector field (or each of their
vector fields) is expressed in the same way in terms of $a$, $b$
and two functions and the quantity $\lambda=\xi+e^{-2\phi}$ is the
same function of $a$, $b$ and $d$, where $d$ is a real or complex
constant.

\section{The Metric}

In a model with the action
\begin{equation}
S = \int d^{4}x \sqrt{-g}
\{R-2\partial_{\mu}\phi\partial^{\mu}\phi -
e^{-2\phi}F_{\mu\nu}F^{\mu\nu} \} \label{2-2}
\end{equation}
where $R$ is the Ricci scalar and $\phi$ and  $F_{\mu\nu}$ are the
dilaton and a  $U(1)$ vector field respectively, we found the
solution ( Eqs (54)-(57) of Ref. \cite{Ky3} for $\psi=2\phi$,
$\alpha=b$ and $\psi_{0}=0$)
\begin{equation}
ds^{2}=-\frac{(r+A)(r+B)}{r(r+b)}dt^{2}
+\frac{r(r+b)}{(r+A)(r+B)}dr^{2} +r(r+b)(d\theta^{2}+sin^{2}\theta
{d{\phi}^{2}})
 \label{2-2}
\end{equation}
\begin{equation}
e^{2\phi}=1+\frac{b}{r},\>\>\> F=\frac{\sqrt{AB}}{\sqrt{2}r^2}dr
\wedge dt +\frac{\sqrt{(b-A)(b-B)}}{\sqrt{2}} sin\theta d\theta
\wedge d\phi
 \label{2-3}
\end{equation}
and in a model with the action
\begin{equation}
S = \int d^{4}x \sqrt{-g}
\{R-2\partial_{\mu}\phi\partial^{\mu}\phi - (g_1e^{2\phi}
+g_2e^{-2\phi}) \} \label{2-4}
\end{equation}
we found a solution ( Eqs (2) and (12)-(14) of Ref  \cite{Ky1} for
 $\psi=2\phi$, $\alpha=b$ and $\psi_{0}=0$) with the same $ds^2$
and
\begin{equation}
e^{-2\phi}=1+\frac{b}{r},\>\>\> F=Q dr \wedge dt,\>\>\>
g_1=\frac{AB}{2Q^2},\>\>\> g_2=\frac{(b-A)(b-B)}{2Q^2}
 \label{2-5}
\end{equation}
where $A, B$ and $b$ are arbitrary real constants. We shall apply
the method of Newman and Janis \cite{Ne1}, \cite{Ra1}to the metric
of expression (\ref{2-2}).

Following this method we replace $dt$ in (\ref{2-2}) by $du$,
where
\begin{equation}
dt=du+\frac{r(r+b)}{(r+A)(r+B)}dr \label{2-6}
\end{equation}
Then Eq. (\ref{2-2}) becomes
\begin{equation}
ds^2=-\frac{(1+\frac{A}{r})(1+\frac{B}{r})}{1+\frac{b}{r}}du^2-2du
dr +r^2(1+\frac{b}{r})(d\theta^2+\sin^2\theta d\phi^2)
 \label{2-7}
\end{equation}
If $g^{\mu\nu}$ is the inverse of the metric of the line element
(\ref{2-7}) and introduce the vectors
\begin{equation}
l^{\mu}=\delta_{1}^{\mu}, \>\>\>
n^{\mu}=\delta_{0}^{\mu}-\frac{(1+\frac{A}{r})(1+\frac{B}{r})}
{2(1+\frac{b}{r})}\delta_{1}^{\mu},\>\>\>
m^{\mu}=\frac{1}{r\sqrt{2(1+\frac{b}{r})}}(\delta_{2}^{\mu}+
\frac{i}{\sin\theta}\delta_{3}^{\mu})
 \label{2-8}
\end{equation}
we get
\begin{equation}
 g^{\mu\nu}=-l^{\mu}n^{\nu}-l^{\nu}n^{\mu}
+m^{\mu}{\bar{m}}^{\nu} +m^{\nu}{\bar{m}}^{\mu} \label{2-9}
\end{equation}
We allow now $r$ to become complex and keeping the same symbols
for the vectors we write
\[l^{\mu}=\delta_{1}^{\mu},
\>\>\>
n^{\mu}=\delta_{0}^{\mu}-\frac{1+\frac{A+B}{2}(\frac{1}{r}+\frac{1}{\bar{r}})+
\frac{AB}{r\bar{r}}}
{2(1+\frac{b}{2}(\frac{1}{r}+\frac{1}{\bar{r}}))}\delta_{1}^{\mu},\>\>\>
m^{\mu}=\frac{1}{\bar{r}\sqrt{2(1+\frac{b}{2}(\frac{1}{r}
+\frac{1}{\bar{r}}))}}(\delta_{2}^{\mu}\]
\begin{equation}
+\frac{i}{\sin\theta}\delta_{3}^{\mu})
 \label{2-10}
\end{equation}
and subsequently we apply to them the complex coordinate
transformation
\begin{equation}
r \rightarrow r'=r +ia\cos\theta,\>\>\> u \rightarrow u'=
u-ia\cos\theta  \label{2-11}
\end{equation}
This transformation gives the $l'^\mu$, $n'^\mu$ and $m'^\mu$
vectors
\begin{equation}
l'^{\mu}=\delta_1^{\mu}, \>\>\>
n'^{\mu}=\delta_0^{\mu}-\frac{\Delta'}{2\rho'^2}\delta_1^\mu,\>\>\>
m'^\mu=\frac{\sqrt{r}}{\sqrt{2\bar{r}\rho'^2}}\{i
a\sin\theta(\delta_0^\mu-\delta_1^\mu)+\delta_2^\mu
+\frac{i}{\sin\theta}\delta_3^\mu\} \label{2-12}
\end{equation}
where
\begin{equation}
\rho'^2=r'(r'+b)+a^2\cos^2\theta,\>\>\>\>\>\>
\Delta'=(r'+A)(r'+B)+a^2\cos^2\theta\label{2-13}
\end{equation}
For the metric $g'^{\mu\nu}$ defined by
\begin{equation}
g'^{\mu\nu}=-l'^\mu n'^\nu-l'^\nu n'^\mu+m'^\mu
{\bar{m'}}^\nu+m'^\nu {\bar{m'}}^\mu \label{2-14}
\end{equation}
we get
\begin{equation}
 g'^{\mu\nu}=\frac{1}{\rho'^2}\left(\begin{array}{cccc}
 a^2\sin^2\theta&-\rho'^2-a^2\sin^2\theta&0&a\\
-\rho'^2-a^2\sin^2\theta&\Delta'+a^2\sin^2\theta&0&-a\\
0&0&1&0\\
a&-a&0&\frac{1}{\sin^2\theta}
\end{array} \right) \label{2-15}
\end{equation}
from which we obtain its inverse
\begin{equation}
 g'_{\mu\nu}=\left(\begin{array}{cccc}
 -\frac{\Delta'}{\rho'^2}&-1&0&\frac{a\sin^2\theta(\Delta'-\rho'^2)}{\rho'^2}\\
-1&0&0&a\sin^2\theta\\
0&0&\rho'^2&0\\
\frac{a\sin^2\theta(\Delta'-\rho'^2)}{\rho'^2}&a\sin^2\theta&0&\sin^2\theta(\rho'^2
+2a^2\sin^2\theta)-\frac{a^2\Delta'\sin^4\theta}{\rho"2}
\end{array} \right) \label{2-16}
\end{equation}
If in the $ds'^2$ coming from the above $g'_{\mu\nu}$ we make the
transformation
\begin{equation}
du'=dt'-\frac{\rho'^2+a^2\sin^2\theta}{\Delta'+a^2\sin^2\theta}dr',\>\>\>\>
d\phi=d\phi'-\frac{a}{\Delta'+a^2\sin^2\theta}dr'
 \label{2-17}
\end{equation}
the only non-diagonal term which remains is $dt'd\phi'$ and we get
if we drop the primes
\begin{equation}
 g_{\mu\nu}=\left(\begin{array}{cccc}
-\frac{\Delta}{\rho^2}&0&0&\frac{a(\Delta-\rho^2)\sin^2\theta}{\rho^2}\\
0&\frac{\rho^2}{\Delta+a^2\sin^2\theta}&0&0\\
0&0&\rho^2&0\\
\frac{a(\Delta-\rho^2)\sin^2\theta}{\rho^2}&0&0&\sin^2\theta(\rho^2+2a^2\sin^2\theta)
-\frac{a^2\Delta\sin^4\theta}{\rho^2} \end{array} \right)
\label{2-18}
\end{equation}
where
\begin{equation}
\rho^2=r(r+b)+a^2\cos^2\theta,\>\>\Delta=(r+A)(r+B)+a^2\cos^2\theta\label{2-19}
\end{equation}
The $g_{\mu\nu}$ of expression (\ref{2-18}) gives
\begin{equation}
|g_{\mu\nu}|=-\rho^4\sin^2\theta \label{2-20}
\end{equation}
\[ds^2=g_{\mu\nu}dx^{\mu}dx^{\nu}=-\frac{\Delta}{\rho^2}dt^2+
\frac{\rho^2}{\Delta+a^2\sin^2\theta}dr^2+ \rho^2d\theta^2  +
\frac{2a(\Delta-\rho^2)\sin^2\theta}{\rho^2}dtd\phi
\]
\begin{equation}
+\{\sin^2\theta(\rho^2+2a^2\sin^2\theta)
-\frac{a^2\Delta\sin^4\theta}{\rho^2}\}d\phi^2
  \label{2-21}
\end{equation}
Also from the $g_{\mu\nu}$ of expression (\ref{2-18}) we get for
any $\Delta$ and $\rho^2$
\begin{equation}
 g^{\mu\nu}=\left(\begin{array}{cccc}
-\frac{\rho^2+2a^2\sin^2\theta
-\frac{a^2\Delta\sin^2\theta}{\rho^2}}{\Delta+a^2\sin^2\theta}&0&0
&\frac{a(\Delta-\rho^2)}{\rho^2(\Delta+a^2\sin^2\theta)}\\
0&\frac{\Delta+a^2\sin^2\theta}{\rho^2}&0&0\\
0&0&\frac{1}{\rho^2}&0\\
\frac{a(\Delta-\rho^2)}{\rho^2(\Delta+a^2\sin^2\theta)}&0&0
&\frac{\Delta}{\rho^2(\Delta+a^2\sin^2\theta)\sin^2\theta}
\end{array}\right)
\label{2-22}
\end{equation}
If we introduce the notation
\begin{equation}
(b-A)(b-B)=2Q_E^2, \>\>\> AB=2Q_M^2, \>\>\> b-A-B=2M \label{2-23}
\end{equation}
we get \cite{To1}
\begin{equation}
b=\frac{Q_E^2-Q_M^2}{M}, \>\>\>
\rho^2=r(r+\frac{Q_E^2-Q_M^2}{M})+a^2\cos^2\theta \label{2-24}
\end{equation}
\begin{equation}
\Delta=r(r+\frac{Q_E^2-Q_M^2}{M})-2Mr+2Q_M^2+a^2\cos^2\theta
=\rho^2-2Mr+2Q_M^2 \label{2-25}
\end{equation}
We shall find two  solutions whose metric is given by Eqs
(\ref{2-18}), (\ref{2-24}) and (\ref{2-25}). This metric has four
arbitrary parameters namely $ M,\> a,\> Q_E $ and $ Q_M $.
Generally all solutions we shall consider in this paper have a
metric $g_{\mu\nu}$ of the form of  Eq (\ref{2-18}). This metric
is a function of the rotation parameter $a$ and of $\rho^2$ and
$\Delta$, which are given by the relations
\begin{equation}
\rho^2=r(r+b)+a^2\cos^2\theta,
\>\>\>\>\>\Delta=r(r+b)-2Mr+c+a^2\cos^2\theta \label{2-25'}
\end{equation}
where $M$, $b$ and $c$ are constants. It is obvious that a metric
of the form of Eqs (\ref{2-18}), (\ref{2-24}) and (\ref{2-25}) is
also of the form of Eqs (\ref{2-18}) and (\ref{2-25'}). However
there are solutions with $\rho^2$ and $\Delta$ of the form of Eq
(\ref{2-25'}) but not of the form of Eqs (\ref{2-24}) and
(\ref{2-25}), that is with $b\neq\frac{Q_E^2-Q_M^2}{M}$ and or
$c\neq2Q_M^2$. Such are the Kerr-Newman solution and the
Reissner-Nordstr\"{o}m solution for electric charge $Q_E$ and
magnetic charge $Q_M$ with $Q_E\neq Q_M $. In these two solutions
we have $b=0$
 and $c=Q_E^2+Q_M^2$.

\section{Model with two Vector Fields}

Consider a model which has besides gravity an axion field $\xi$, a
dilaton field $\phi$, two vector fields $F_{\mu\nu}=\partial_\mu
A_\nu-{\partial}_\nu A_\mu$ and $F'_{\mu\nu}=\partial_\mu
A'_\nu-{\partial}_\nu A'_\mu$ and action
\[S = \int d^{4}x \sqrt{-g}
\{R-2\partial_{\mu}\phi\partial^{\mu}\phi-
\frac{1}{2}e^{4\phi}\partial_{\mu}\xi\partial^{\mu}\xi-
e^{-2\phi}(F_{\mu\nu}F^{\mu\nu}+F'_{\mu\nu}{F'}^{\mu\nu})\]
\begin{equation}
 -\xi(F_{\mu\nu}{F^\ast}^{\mu\nu}+F'_{\mu\nu}{F'^\ast}^{\mu\nu})\}
\label{3-1}
\end{equation}
where
\begin{equation}
{F^{\ast}}^{\mu\nu}=\frac{1}{2\sqrt{-g}}\epsilon^{\mu\nu\sigma\tau}F_{\sigma
\tau},
\>\>\>{F'^{\ast}}^{\mu\nu}=\frac{1}{2\sqrt{-g}}\epsilon^{\mu\nu\sigma\tau}F'_{\sigma
\tau}, \>\>\> \epsilon^{0123}=+1 \label{3-2}
\end{equation}
Also we define $\lambda,\>\>{F_{\pm}}^{\mu\nu}$  and
${F'_{\pm}}^{\mu\nu}$  by
\begin{equation}
\lambda=\xi+ie^{-2\phi},\>\> {F_{\pm}}^{\mu\nu}=F^{\mu\nu}\pm
i{F^\ast}_{\mu\nu}, \>\> {F'_{\pm}}^{\mu\nu}=F'^{\mu\nu}\pm
i{F'^\ast}^{\mu\nu } \label{3-3}
\end{equation}
Then the action can be written in the form \cite{Sh1}
\begin{equation}
S = \int d^{4}x \sqrt{-g}
\{R-\frac{1}{2}e^{4\phi}\partial_{\mu}\lambda\partial^\mu\bar{\lambda}
+\frac{i}{4}(\lambda F_{+}^2-\bar{\lambda}F_{-}^2)
+\frac{i}{4}(\lambda {F'_{+}}^2-\bar{\lambda}{F'_{-}}^2)\}
\label{3-4}
\end{equation}
and we get the equations of motion
\begin{equation}
R_{\mu\nu}=\frac{1}{4}e^{4\phi}(\partial_\mu\bar\lambda
\partial_\nu\lambda +\partial_\nu\bar\lambda\partial_\mu\lambda)
+2e^{-2\phi}(F_{\mu\sigma}F_\nu^\sigma-\frac{g_{\mu\nu}}{4}F^2)
+2e^{-2\phi}(F'_{\mu\sigma}{F'_\nu}^\sigma-\frac{g_{\mu\nu}}{4}F'^2)
\label{3-5}
\end{equation}
\begin{equation}
\nabla_\mu(\lambda F_{+}^{\mu\nu}-\bar{\lambda}F_{-}^{\mu\nu})=0,
\>\>\> \nabla_\mu(\lambda
{F'_{+}}^{\mu\nu}-\bar{\lambda}{F'_{-}}^{\mu\nu})=0 \label{3-6}
\end{equation}
\begin{equation}
e^{4\phi}\nabla_\mu\partial^\mu\lambda+ie^{6\phi}\partial
_\mu\lambda\partial^\mu\lambda
-\frac{i}{2}({F_{-}}^2+{F'_{-}}^2)=0 \label{3-7}
\end{equation}
To find a solution of the above equations with the metric of Eqs
(\ref{2-18}), (\ref{2-24}) and  (\ref{2-25}) we shall make the
following ansatz for the fields $F_{\mu\nu}$ and $F'_{\mu\nu}$
\[F_{rt}=\zeta,\>\>\> F_{r\phi}=-a\sin^2\theta\zeta, \>\>\>
F_{\theta t}=-a\sin\theta\eta, \>\>\>
F_{\theta\phi}=\{r(r+b)+a^2\}\sin\theta\eta\]
\begin{equation}
F_{r\theta}=F_{t\phi}=0  \label{3-8}
\end{equation}
\[F'_{rt}=\zeta',\>\>\> F'_{r\phi}=-a\sin^2\theta\zeta', \>\>\>
F'_{\theta t}=-a\sin\theta\eta', \>\>\>
F'_{\theta\phi}=\{r(r+b)+a^2\}\sin\theta\eta'\]
\begin{equation}
F'_{r\theta}=F'_{t\phi}=0  \label{3-9}
\end{equation}
The previous expressions give for the metric of Eqs (\ref{2-18}),
(\ref{2-24}) and (\ref{2-25})
\begin{equation}
F^2=2(\eta^2-\zeta^2), \>\>\> FF^\ast=-4\zeta\eta, \>\>\>
F'^2=2(\eta'^2-\zeta'^2), \>\>\> F'F'^\ast=-4\zeta'\eta'
\label{3-10}
\end{equation}
Also we have ${F^\ast}^2=-F^2$ and ${{F'}^\ast}^2=-F'^2$ Generally
the vector fields $F_{\mu\nu}$ of all solutions we shall consider
in this paper are of the above form, which means that each
$F_{\mu\nu}$ is expressed in terms of the rotation parameter $a$,
a constant $b$ and two functions.

\section{Reduction of the Metric Equations}

For the metric of Eqs (\ref{2-18}), (\ref{2-24}) and (\ref{2-25})
we have the following non zero components of the Ricci tensor
\cite{Bo1}
\begin{equation}
R_{tt}=\frac{(\Delta+2a^2\sin^2\theta)K}{\rho^8},\>\>\>\>\>
R_{rr}=\frac{\rho^2(b^2-2(Q_E^2+Q_M^2))+a^2b^2\sin^2\theta}{2(\Delta
+a^2\sin^2\theta)\rho^4} \label{4-1}
\end{equation}
\begin{equation}
R_{\theta \theta} = \frac{2K+a^2b^2\sin^2\theta}{2\rho^4},
\>\>\>\>\>\>\>
R_{t\phi}=-\frac{a\sin^2\theta(\Delta+\rho^2+2a^2\sin^2\theta)K}{\rho^8}
\label{4-2}
\end{equation}
\begin{equation}
R_{\phi\phi}=\frac{\sin^2\theta\{(\rho^2+a^2\sin^2\theta)^2+(\Delta
+a^2\sin^2\theta
)a^2\sin^2\theta\}K}{\rho^8} \label{4-3}
\end{equation}
where
\begin{equation}
K=Q_E^2(r^2+a^2\cos^2\theta)+Q_M^2\{(r+b)^2+a^2\cos^2\theta\}
\label{4-4}
\end{equation}
Also if the components of $F_{\mu\nu}$ are given by Eqs
(\ref{3-8}) we get
\begin{equation}
F_{t\mu}F_t^\mu-\frac{g_{tt}}{4}F^2
=\frac{(\Delta+2a^2\sin^2\theta)}{2\rho^2}(\zeta^2
+\eta^2),\>\>\>\>F_{\theta\mu}F_\theta^\mu-\frac{g_{\theta\theta}}{4}F^2
=\frac{\rho^2(\zeta^2+\eta^2)}{2} \label{4-5}
\end{equation}
\begin{equation}
 F_{r\mu}F_r^\mu-\frac{g_{rr}}{4}F^2
=-\frac{\rho^2(\zeta^2+\eta^2)}{2(\Delta+a^2\sin^2\theta)}
\label{4-6}
\end{equation}
\begin{equation}
F_{t\mu}F_\phi^\mu-\frac{g_{t\phi}}{4}F^2=-\frac{a\sin^2\theta(\Delta+\rho^2
+2a^2\sin^2\theta)(\zeta^2+\eta^2)}{2\rho^2} \label{4-7}
\end{equation}
\begin{equation}
F_{\phi\mu}F_\phi^\mu-\frac{g_{\phi\phi}}{4}F^2
=\frac{\sin^2\theta(a^2\Delta\sin^2\theta
+(\rho^2+a^2\sin^2\theta)^2+a^4\sin^4\theta)(\zeta^2+\eta^2)}{2\rho^2}
\label{4-8}
\end{equation}
\begin{equation}
F_{\mu\sigma}F_\nu^\sigma-\frac{g_{\mu\nu}}{4}F^2=0\>\>\>\>\>\mbox{for}\>\>(\mu,
\nu)=(r,\theta),\>(t,r),\>(t,\theta),\>(\phi,r),\>(\phi,\theta)
\label{4-9}
\end{equation}
The expressions
$F'_{\mu\sigma}{F'_\nu}^\sigma-\frac{g_{\mu\nu}}{4}F'^2$ are
obtained from the above expressions if we replace $\zeta^2+\eta^2$
by $\zeta'^2+\eta'^2$.Also in the paper we shall assume that
$\lambda$, $F_{\mu\nu}$ and $F'_{\mu\nu}$ do not depend on $t$ and
$\phi$ and we shall define $y$ by the relation
\begin{equation}
  y=a\cos\theta \label{4-10}
\end{equation}
Then Eq. (\ref{3-5}) gives
\begin{equation}
\zeta^2+\eta^2+\zeta'^2+\eta'^2=\frac{Ke^{2\phi}}{\rho^6},\>\>\mbox{for}
\>\>\>(\mu,\nu)=(t,t),\>\> (\phi,\phi),\>\> (t,\phi) \label{4-11}
\end{equation}
\begin{equation}
\partial_r \lambda\partial_r \bar{\lambda}=\frac{b^2e^{-4\phi}}{\rho^4},
\>\>\mbox{for}\>\>\>(\mu,\nu)=(r,r) \label{4-12}
\end{equation}
\begin{equation}
\partial_y \lambda\partial_y \bar{\lambda}=\frac{b^2e^{-4\phi}}{\rho^4},
\>\>\mbox{for}\>\>\>(\mu,\nu)=(\theta,\theta) \label{4-13}
\end{equation}
\begin{equation}
\partial_r \lambda\partial_y \bar{\lambda}+\partial_y
\lambda\partial_r\bar{\lambda}=0,
\>\>\mbox{for}\>\>\>(\mu,\nu)=(r,\theta) \label{4-14}
\end{equation}
while it is identically satisfied for $(\mu,\nu)=(t,r)$,
$(t,\theta)$, $(\phi,r)$ and $(\phi,\theta)$. To derive Eqs
(\ref{4-12}) and (\ref{4-13}) we used  Eq. (\ref{4-11}).
Multiplying Eqs (\ref{4-12}) and (\ref{4-13}) by parts and using
Eq. (\ref{4-14}) we get
\begin{equation}
\partial_r \lambda\partial_r \bar{\lambda}\partial_y \lambda\partial_y \bar{\lambda}
=\frac{b^4e^{-8\phi}}{\rho^8}=-(\partial_y
\lambda)^2(\partial_r\bar{\lambda})^2\label{4-15}
\end{equation}
From Eqs (\ref{4-12}) and (\ref{4-15}) we get
\begin{equation}
\partial_y\lambda)\partial_r\bar{\lambda}=\pm
i\frac{b^2e^{-4\phi}}{\rho^4}=\pm
i(\partial_r\lambda)(\partial_r\bar{\lambda}) \label{4-16}
\end{equation}
Therefore we get
\begin{equation}
\partial_y\lambda=\pm i\partial_r\lambda
\label{4-17}
\end{equation}
whose solution is $\lambda=\lambda(r+iy)$ for the + sign and
$\lambda=\lambda(r-iy)$ for the - sign. We shall take
\begin{equation}
\lambda=\lambda(r+iy) \label{4-18}
\end{equation}
where $\lambda(r+iy)$ is an arbitrary function of its arguments.
If relations (\ref{4-12}) and (\ref{4-18}) hold Eqs (\ref{4-13})
and (\ref{4-14}) are satisfied. Therefore if Eqs (\ref{4-11}),
(\ref{4-12}) and (\ref{4-18}) hold Eq. (\ref{3-5}) is satisfied
for all values of $\mu$ and $\nu$. In other words all metric
equations are reduced to Eqs (\ref{4-11}), (\ref{4-12}) and
(\ref{4-18}), where $y$ is given by Eq. (\ref{4-10}).

\section{Solution of the Vector Field Equations}

We have
\begin{equation}
\lambda {F_+}^{\mu\nu}-\bar{\lambda}{F_-}^{\mu\nu} = 2
i(e^{-2\phi}F^{\mu\nu} +\xi{F^\ast}^{\mu\nu}) \label{5-1}
\end{equation}
and the first of Eqs (\ref{3-6}) becomes
\begin{equation}
\partial_\mu(\sqrt{-g}(e^{-2\phi}F^{\mu\nu}+\xi{F^\ast}^{\mu\nu}))=0
\label{5-2}
\end{equation}
Then since $\lambda$, $F_{\mu\nu}$ and $g$ are not functions of
$t$ and $\phi$ while $F_{r\theta}=F_{t\phi}=0$, the above relation
is satisfied for $\nu=r $ and $\nu=\theta $. Eq. (\ref{5-2}) for
the metric of Eqs (\ref{2-18}), (\ref{2-24}) and (\ref{2-25})
gives
\begin{equation}
\mbox{for}\; \nu=t :\>\>
\partial_r\{(\rho^2+a^2\sin^2\theta)(e^{-2\phi}\zeta+\xi\eta)\}
+\partial_y\{a^2\sin^2\theta(e^{-2\phi}\eta-\xi\zeta)\}=0
\label{5-3}
\end{equation}
\begin{equation}
\mbox{and for}\>\>\>\>\>\>\> \nu=\phi
:\>\>\>\partial_r(e^{-2\phi}\zeta+\xi\eta)+\partial_y(e^{-2\phi}\eta-\xi\zeta)=0
\label{5-4}
\end{equation}
Multiplying Eq. (\ref{5-4}) by $a^2$ and subtracting from Eq.
(\ref{5-3}) we get
\begin{equation}
\partial_r\{r(r+b)(e^{-2\phi}\zeta+\xi\eta)\}-\partial_y\{y^2(e^{-2\phi}\eta
-\xi\zeta)\}=0 \label{5-5}
\end{equation}
The general solution of Eq. (\ref{5-4}) is
\begin{equation}
e^{-2\phi}\zeta+\xi\eta=\partial_yP(r,y),\>\>\>\>\>\>\>e^{-2\phi}\eta
-\xi\zeta=-\partial_rP(r,y) \label{5-6}
\end{equation}
where $P(r,y)$ is an arbitrary function of its arguments. Then
substituting the expressions (\ref{5-6}) in (\ref{5-5})we get
\begin{equation}
\partial_r\partial_y\{(r(r+b)+y^2)P\}=0
\label{5-7}
\end{equation}
Therefore we get
\begin{equation}
P=\frac{G(r)+H(y)}{r(r+b)+y^2}=\frac{G(r)+H(y)}{\rho^2}
\label{5-8}
\end{equation}
where $G(r)$ and $H(y)$ are arbitrary functions of their arguments
and Eq. (\ref{2-24}) was used. If we  solve Eqs (\ref{5-6}) for
$\zeta$ and $\eta$ we get
\begin{equation}
\zeta=(\lambda\bar{\lambda})^{-1}(\xi\partial_rP
+e^{-2\phi}\partial_yP),\>\>\>\>\>\eta=
(\lambda\bar{\lambda})^{-1}(\xi\partial_yP -e^{-2\phi}\partial_rP)
\label{5-9}
\end{equation}

Proceeding in an analogous fashion we can solve the second of Eqs
(\ref{3-6}) and find $P'$, $\zeta'$ and $\eta'$. We get
\begin{equation}
P'=\frac{G'(r)+H'(y)}{\rho^2} \label{5-10}
\end{equation}
\begin{equation}
\zeta'=(\lambda\bar{\lambda})^{-1}(\xi\partial_rP'
+e^{-2\phi}\partial_yP'),\>\>\>\>\>\eta'=
(\lambda\bar{\lambda})^{-1}(\xi\partial_yP'
-e^{-2\phi}\partial_rP') \label{5-11}
\end{equation}
with $G'(r)$ and $H'(y)$ arbitrary functions of their arguments.
From Eqs (\ref{5-9}) and (\ref{5-11}) we get
\begin{equation}
\zeta^2+\eta^2+\zeta'^2+\eta'^2=\frac{1}{\lambda\bar{\lambda}}((\partial_rP)^2
+(\partial_yP)^2+ (\partial_rP')^2+ (\partial_yP')^2)\label{5-12}
\end{equation}
and Eq. (\ref{4-11}) becomes
\begin{equation}
\frac{Ke^{2\phi}\lambda\bar{\lambda}}{\rho^6}=(\partial_rP)^2+(\partial_yP)^2+
(\partial_rP')^2+ (\partial_yP')^2 \label{5-13}
\end{equation}

\section{Solution of the Axion-Dilaton Equation}

We shall solve now the axion-dilaton Eq. (\ref{3-7}). From Eqs
(\ref{3-3}), (\ref{3-10}), (\ref{5-9}) and (\ref{5-11}) we get
\begin{equation}
F_{-}^2=2(F^2-iFF^\ast)=-4(\zeta-i\eta)^2=-4{\bar{\lambda}}^{-2}(\partial_rP
-i\partial_yP)^2   \label{6-1}
\end{equation}
\begin{equation}
{F'}_{-}^2=2({F'}^2-iF'{F'}^\ast)=-4(\zeta'-i\eta')^2
=-4{\bar{\lambda}}^{-2}(\partial_rP' -i\partial_yP')^2 \label{6-2}
\end{equation}
and Eq. (\ref{3-7}) becomes
\begin{equation}
\nabla_\mu\partial^\mu\lambda+ie^{2\phi}\partial_\mu\lambda\partial^\mu\lambda
+2ie^{-4\phi}{\bar{\lambda}}^{-2}\{(\partial_rP
-i\partial_yP)^2+(\partial_rP' -i\partial_yP')^2 \}=0
 \label{6-3}
\end{equation}
Since for the metric of Eqs (\ref{2-18}), (\ref{2-24}) and
(\ref{2-25}) and for $\lambda=\lambda(r+iy)$ we get
\begin{equation}
\partial_\mu\lambda\partial^\mu\lambda=\frac{\Delta}{\rho^2}(\partial_r\lambda)^2
\label{6-4}
\end{equation}
\begin{equation}
\nabla_\mu\partial^\mu\lambda=\frac{\triangle}{\rho^2}\partial_r\partial_r\lambda
+\frac{1}{\rho^2}\{2(r-iy)+b+\frac{2}{b}(Q_M^2-Q_E^2)\}\partial_r\lambda
\label{6-5}
\end{equation}
Eq. (\ref{6-3}) becomes
\[\Delta\partial_r\partial_r\lambda+\{2(r-iy)+b+\frac{2}{b}(Q_M^2
-Q_E^2)\}\partial_r\lambda +i\Delta
e^{2\phi}(\partial_r\lambda)^2\]
\begin{equation}
+2i\rho^2e^{-4\phi}{\bar{\lambda}}^{-2}\{(\partial_rP-i\partial_yP)^2
+(\partial_rP'-i\partial_yP')^2\}\equiv\Lambda=0 \label{6-6}
\end{equation}

To find a solution of Eq. (\ref{6-6}) we multiply this equation by
$\partial_r\bar{\lambda}$, then we multiply its complex conjugate
by $\partial_r\lambda$ and subsequently we sum the resulting
expressions by parts. We get
$\Lambda\partial_r\bar{\lambda}+\bar{\Lambda}\partial_r\lambda=0$
or if we use Eq. (\ref{4-12})
\[2ib\{Q_M^2((r+b)^2-y^2)-Q_E^2(r^2-y^2)\}=\rho^8\{(\partial_rP
-i\partial_yP)^2\partial_r({\bar{\lambda}}^{-1})\]
\begin{equation}
-(\partial_rP
+i\partial_yP)^2\partial_r(\lambda^{-1})\}+\rho^8\{(\partial_rP'
-i\partial_yP')^2\partial_r({\bar{\lambda}}^{-1})-(\partial_rP'
+i\partial_yP')^2\partial_r(\lambda^{-1})\}\label{6-7}
\end{equation}
Eq.(\ref{6-7}) is satisfied if $\lambda$, $P$ and $P'$ satisfy the
relations
\begin{equation}
-2ibQ_E^2(r^2-y^2)=\rho^8\{(\partial_rP
-i\partial_yP)^2\partial_r({\bar{\lambda}}^{-1}) -(\partial_rP
+i\partial_yP)^2\partial_r(\lambda^{-1})\}\label{6-8}
\end{equation}
\begin{equation}
2ibQ_M^2((r+b)^2-y^2)=\rho^8\{(\partial_rP'
-i\partial_yP')^2\partial_r({\bar{\lambda}}^{-1})-(\partial_rP'
+i\partial_yP')^2\partial_r(\lambda^{-1})\}\label{6-9}
\end{equation}
Of course $\lambda$, $P$ and $P'$ must be of the form given by Eqs
(\ref{4-18}), (\ref{5-8}) and (\ref{5-10}) respectively and they
must be a solution of Eq. (\ref{6-6}). All these requirements are
satisfied if
\begin{equation}
\lambda=\frac{i(r+iy+b)}{r+iy} \label{6-10}
\end{equation}
\begin{equation}
P=\frac{Q_Ey}{\rho^2}, \>\>\>P'=\frac{Q_M(r+b)}{\rho^2}
\label{6-11}
\end{equation}
From Eq. (\ref{6-10}) we get
\begin{equation}
\xi=\frac{by}{r^2+y^2}, \>\>\>\>e^{-2\phi}=\frac{\rho^2}{r^2+y^2}
\label{6-12}
\end{equation}
and Eq. (\ref{4-12}) is satisfied. Also from Eqs (\ref{5-9}),
(\ref{5-11}) and (\ref{6-10})-(\ref{6-12}) we get
\begin{equation}
\zeta=\frac{Q_E(r^2-y^2)}{\rho^4},\>\>\>\>\eta=\frac{2Q_Ery}{\rho^4},\>\>\>\>
\zeta'=-\frac{Q_My(2r+b)}{\rho^4},\>\>\>\>\eta'=\frac{Q_M\{r(r+b)-y^2\}}{\rho^4}
\label{6-13}
\end{equation}
and Eq. (\ref{4-11}) is satisfied, which means that all equations
are satisfied.

In summary our solution has a metric given by Eqs (\ref{2-18}),
(\ref{2-24}) and (\ref{2-25}), $\lambda$, $\xi$ and $e^{-2\phi}$
given by Eqs (\ref{4-10}), (\ref{6-10}) and (\ref{6-12}) or by the
relations
\begin{equation}
\lambda=\frac{i(r+ia\cos\theta+b)}{r+ia\cos\theta},\>\>\xi=\frac{ab\cos\theta}{r^2
+a^2\cos^2\theta},\>\>\>\>e^{-2\phi}=\frac{\rho^2}{r^2+a^2\cos^2\theta}
\label{6-14}
\end{equation}
and $F_{\mu\nu}$ and $F'_{\mu\nu}$ given by Eqs (\ref{3-8}) and
(\ref{3-9}) with $\zeta$, $\eta$, $\zeta'$ and $\eta'$ given by
Eqs. (\ref{4-10}) and (\ref{6-13}) or by the relations
\[\zeta=\frac{Q_E(r^2-a^2\cos^2\theta)}{\rho^4},\>\>\>\>
\eta=\frac{2Q_Ear\cos\theta}{\rho^4},\>\>\>\>
\zeta'=-\frac{Q_Ma\cos\theta(2r+b)}{\rho^4}\]
\begin{equation}
\eta'=\frac{Q_M\{r(r+b)-a^2\cos^2\theta\}}{\rho^4} \label{6-15}
\end{equation}

The solution we have found has $\xi$ and $\phi$ fields with zero
asymptotic values. However we can easily construct a solution with
arbitrary asymptotic values of $\xi$ and $\phi$. Indeed we find
using Eqs (\ref{3-5})-(\ref{3-7}) that the expressions
\begin{equation}
\tilde{g_{\mu\nu}}=g_{\mu\nu},\>\>
\tilde{\xi}=e^{-2\phi_\infty}\xi+\xi_\infty,\>\>e^{-2\tilde{\phi}}
=e^{-2\phi_\infty} e^{-2\phi}, \>\>
\tilde{F_{\mu\nu}}=e^{\phi_\infty}F_{\mu\nu},\>\>
\tilde{F'_{\mu\nu}}=e^{\phi_\infty}F'_{\mu\nu} \label{6-16}
\end{equation}
where $g_{\mu\nu}$, $\xi$, $e^{-2\phi}$, $F_{\mu\nu}$,and
$F'_{\mu\nu}$ are given by Eqs (\ref{2-18}), (\ref{2-24}),
(\ref{2-25}) (\ref{3-8}), (\ref{3-9}), (\ref{6-14}) and
(\ref{6-15}) and $\xi_\infty$ and $\phi_\infty$ are arbitrary real
constants, give a solution with arbitrary asymptotic values
$\xi_\infty$ and $\phi_\infty$ of $\xi$ and $\phi$.

\section{Physical Properties of the Solution}

We shall describe now some physical properties of our solution. We
get for large $r$
\begin{equation}
g_{tt}=-\frac{\Delta}{\rho^2}\sim-(1-\frac{2M}{r})+O(\frac{1}{r^2})
\label{7-1}
\end{equation}
Therefore the solution is asymptotically flat and the parameter
$M$ is its mass. The $F_{\mu\nu}$ of the first vector field can be
obtained from the vector potential
\begin{equation}
A_t=-
\frac{Q_Er}{\rho^2},\>\>A_\phi=\frac{Q_Ear\sin^2\theta}{\rho^2},
\>\> A_r=A_\theta=0 \label{7-2}
\end{equation}
and the $F'_{\mu\nu}$ of the second vector field from the vector
potential
\begin{equation}
A'_t=\frac{Q_Ma\cos\theta}{\rho^2}, \>\>
A'_\phi=-\frac{Q_M(\rho^2+a^2\sin^2\theta)\cos\theta}{\rho^2},
\>\>A'_r=A'_\theta=0 \label{7-3}
\end{equation}
Since asymptotically
\[F_{rt}=\frac{Q_E(r^2-a^2\cos^2\theta)}{\rho^4}\sim\frac{Q_E}{r^2}
+O(\frac{1}{r^3})\],
\begin{equation}
F'_{\theta\phi}=\frac{Q_M\{r(r+b)+a^2\}\{r(r+b)
-a^2\cos^2\theta\}\sin\theta}{\rho^4}\sim
Q_M\sin\theta+O(\frac{1}{r}) \label{7-4}
\end{equation}
$Q_E$ is an electric charge and $Q_M$ is a magnetic charge.

The angular momentum $J$ and the magnetic moment $\mu$ of a
solution are obtained from the asymptotic form of $g_{t\phi}$ and
$A_\phi$ respectively by the relations \cite{Ho1}
\begin{equation}
g_{t\phi}\sim-\frac{2J}{r}\sin^2\theta+O(\frac{1}{r^2}),\>\>
A_\phi\sim\frac{\mu\sin^2\theta}{r}+O(\frac{1}{r^2}) \label{7-5}
\end{equation}
Since for our solution we have asymptotically
\begin{equation}
g_{t\phi}=\frac{a(\Delta-\rho^2)\sin^2\theta}{\rho^2}\sim-\frac{2Ma\sin^2\theta}{r}
+O(\frac{1}{r^2})\label{7-6}
\end{equation}
\begin{equation}
A_\phi=\frac{Q_Ear\sin^2\theta}{\rho^2}\sim\frac{Q_Ea\sin^2\theta}{r}
+O(\frac{1}{r^2}) \label{7-7}
\end{equation}
the angular momentum $J$ and the magnetic moment  $\mu$ of our
solution are given by the relations
\begin{equation}
J =Ma, \>\>\>\> \mu=Q_Ea \label{7-8}
\end{equation}
If we define the gyromagnetic ratio $g$ by the relation
$\mu=g\frac{Q_EJ}{2M}$ we get from Eqs (\ref{7-8})
\begin{equation}
g=2 \label{7-9}
\end{equation}

The infinite red shift surfaces occur when \cite{Ad1}
\begin{equation}
g_{tt}=-\frac{\Delta}{\rho^2}=0 \label{7-10}
\end{equation}
From Eqs (\ref{2-25}) and (\ref{7-10}) we find that we have two
infinite red shift surfaces $r^S_\pm$ where
\begin{equation}
r^S_\pm=\frac{1}{2M}\{
2M^2-Q_E^2+Q_M^2\pm(4M^4-4M^2(Q_E^2+Q_M^2)+(Q_E^2-Q_M^2)^2
-4J^2\cos^2\theta)^{\frac{1}{2}}\} \label{7-11}
\end{equation}
Therefore the infinite red shift surfaces are closed axially
symmetric surfaces.

 The event horizons are the surfaces \cite{Ad1}
\begin{equation}
\Delta+a^2\sin^2\theta=0 \label{7-12}
\end{equation}
Taking into account  Eqs (\ref{2-24}), (\ref{2-25}) and
(\ref{7-12}) we find that we have two horizon surfaces $r^H_\pm$
\[r^H_\pm=M-\frac{b}{2} \pm \sqrt{(M-\frac{b}{2}
)^2-(2Q_M^2+a^2)}\]
\begin{equation}
=\frac{1}{2M}\{
2M^2-Q_E^2+Q_M^2\pm(4M^4-4M^2(Q_E^2+Q_M^2)+(Q_E^2-Q_M^2)^2-4J^2)^{\frac{1}{2}}\}
\label{7-13}
\end{equation}
The above expressions are obtained from the expressions $r_\pm^S$
for $\theta=0$ and $\theta=\pi$. The outer and the inner horizons
are closed surfaces, are contained within the corresponding
infinite red shift surfaces and coincide with them at $\theta=0$
and $\theta=\pi$. If we compute the Ricci scalar $R$ and the
curvature scalar $R^2=R_{\mu\nu\sigma\tau}R^{\mu\nu\sigma\tau}$
which are obtained from our metric given by Eqs (\ref{2-18}),
(\ref{2-24}) and (\ref{2-25}) we get \cite{Bo1}
\begin{equation}
R=\frac{b^2(\Delta+2a^2\sin^2\theta)}{2\rho^6} \label{7-14}
\end{equation}
\begin{equation}
R^2 =\frac{Y(r,M,Q_E^2,Q_M^2)} {4\rho^{12}} \label{7-14'}
\end{equation}
where $Y(r,M,Q_E^2,Q_M^2)$ is a complicated polynomial of its
arguments. This means that we have irremovable singularities only
at
\begin{equation}
\rho^2=r(r+\frac{Q_E^2-Q_M^2}{M})+a^2\cos^2\theta=0 \label{7-15}
\end{equation}
Since we shall assume as previously \cite{Ky3} that
$b=\frac{Q_E^2-Q_M^2}{M}>0$ Eq. (\ref{7-15}) for $a\neq0$ is
satisfied if
\begin{equation}
r=\cos\theta=0 \label{7-16}
\end{equation}
Therefore for $a\neq0$ we have a ring singularity, as in the case
of the metric of Kerr \cite{Ra1}. Also from Eqs (\ref{4-10}) and
(\ref{6-12}) we find that for $a\neq0$ the axion and the dilaton
are singular only when $r=\cos\theta=0$. When $a=0$ from Eqs
(\ref{4-10}), (\ref{6-12}) and (\ref{7-14})-(\ref{7-14'}) we find
that the axion field vanish, the dilaton field becomes singular
only at $r=0$ and the metric has irremovable singularity only at
$r=0$ \cite{Ky3}. From Eq. (\ref{7-13}) we find that the horizons
disappear when $4J^2>4M^4-4M^2(Q_E^2+Q_M^2) +(Q_E^2-Q_M^2)^2$ and
therefore the extremal limit corresponds to
\begin{equation}
4J^2\rightarrow 4M^4-4M^2(Q_E^2+Q_M^2)+(Q_E^2-Q_M^2)^2
\label{7-16'}
\end{equation}
To find the area $N$ of the outer horizon we consider the line
element $ds^2$ of Eq. (\ref{2-21}) for $dt=dr=0$ and calculate its
determinant $g'$ at ${r^H}_{+}$. Since
\begin{equation}
g'=\{r_+^H(r_+^H+b)+a^2\}^2\sin^2\theta=
\{r_+^H(r_+^H+\frac{Q_E^2-Q_M^2}{M})+a^2\}^2\sin^2\theta
\label{7-17}
\end{equation}
we get
\[N=\int_{0}^{\pi}d\theta\int_{0}^{2\pi}d\phi\sqrt{g'}=
4\pi\{r_+^H(r_+^H+b)+a^2\}\]
\begin{equation}
=4\pi\{2M^2-Q_E^2-Q_M^2+(4M^4-4M^2(Q_E^2+Q_M^2)+
(Q_E^2-Q_M^2)^2-4J^2)^{\frac{1}{2}}\} \label{7-18}
\end{equation}

The surface gravity $k$ is calculated at the outer horizon $r_+^H$
by the relation \cite{Se1}
\begin{equation}
k=lim_{r \rightarrow r_+^H}\sqrt{g^{rr}}\partial_r
\sqrt{-g_{tt}}|_{\theta=0} \label{7-19}
\end{equation}
We get
\[k=\frac{r_{+}^H-r_{-}^H}{2\{r_+^H(r_+^H+b)+a^2\}}\]
\[=\frac{(4M^4-4M^2(Q_E^2+Q_M^2)+(Q_E^2-Q_M^2)^2-4J^2)^{\frac{1}{2}}}{2M\{2M^2-Q_E^2
-Q_M^2+(4M^4-4M^2(Q_E^2+Q_M^2)+(Q_E^2-Q_M^2)^2-4J^2)^{\frac{1}{2}}
\}}\]
\begin{equation}
=2\pi T_H \label{7-20}
\end{equation}
where $T_H$ is the Hawking temperature. In the extremal limit we
have $k=\frac{1}{4M}$ if $J=Q_M=0$ or $J=Q_E=0$ and $k=0$ in all
other cases. The angular velocity $\Omega$ at the horizon is
calculated from the relation \cite{Se1}
\begin{equation}
g_{tt}+2g_{t\phi}\Omega+g_{\phi\phi}=0 \label{7-21}
\end{equation}
Therefore
\[\Omega=-\frac{g_{t\phi}}{g_{\phi\phi}}|_{r=r_+^H}=\frac{a}{r_+^H(r_+^H+b)+a^2}\]
\begin{equation}
=\frac{J}{M\{2M^2-Q_E^2
-Q_M^2+(4M^4-4M^2(Q_E^2+Q_M^2)+(Q_E^2-Q_M^2)^2-4J^2)^{\frac{1}{2}}
\}} \label{7-22}
\end{equation}
Expressions (\ref{7-13}), (\ref{7-18}), (\ref{7-20}) and
(\ref{7-22}) for $Q_E=Q$ and $Q_M=0$ are identical with the
corresponding expressions of Sen \cite{Se1}.
\vspace{1cm}

\section{Special Cases I}

The solution we have found has the four parameters $M, \>\> a, \>\> Q_E $
and $Q_M$, which can take arbitrary values. All these parameters appear
in the metric and for certain values of them a number of known solutions
or metrics of known solutions can
be obtained. This will be examined in this section. \vspace{3mm}

(1) Solution of Sen with electric charge $Q_E$ only
 \cite{Se1}\vspace{3mm}

If we put in our solution
\begin{equation}
Q_M=0\label{8-1}
\end{equation}
we get from Eqs (\ref{2-24}), (\ref{2-25}) and (\ref{4-10})
\begin{equation}
b=\frac{Q_E^2}{M},\>\> \rho^2=r(r+\frac{Q_E^2}{M})+y^2,\>\>\>
\Delta=r(r+\frac{Q_E^2}{M})-2Mr+y^2
\label{8-2}
\end{equation}
and from Eqs (\ref{6-10}), (\ref{6-12}) and (\ref{6-13})
\begin{equation}
\lambda=\frac{i(r+iy+b)}{r+iy},\>\>\>\>\> \xi=\frac{by}{r^2+y^2},
\>\>\>\>\> e^{-2\phi}=\frac{\rho^2}{r^2+y^2} \label{8-3}
\end{equation}
\begin{equation}
\zeta=\frac{Q_E(r^2-y^2)}{\rho^4},\>\>\>\>\eta=\frac{2Q_Ery}{\rho^4}\label{8-4}
\end{equation}
\begin{equation}
\zeta'=\eta'=0
\label{8-5}
\end{equation}
Eqs (\ref{8-2})-(\ref{8-4}) are identical with Eqs (\ref{B-21})-(\ref{B-23})
 since  $b$ is given by Eq. (\ref{B-20}). Therefore for $Q_M=0$ our
solution becomes the solution of Sen with electric charge only
\cite{Se1}. \vspace{3mm}

(2) Metric of Kerr-Newmam solution with electric charge $Q_E$ and
equal magnetic charge $Q_M$ \cite{Ca1}\vspace{3mm}

If in our solution we put
\begin{equation}
Q_E=Q_M
\label{8-6}
\end{equation}
we get from Eq. (\ref{2-24}) $b=0$ from  Eqs (\ref{2-24}), (\ref{2-25})
and (\ref{4-10})
\begin{equation}
 \rho^2=r^2+y^2,\>\>\>\>\>
\Delta=r^2-2Mr+y^2+2Q_M^2=r^2-2Mr+y^2+Q_E^2+Q_M^2
\label{8-7}
\end{equation}
and from Eqs (\ref{6-10}), (\ref{6-12}) and (\ref{6-13})
\begin{equation}
\xi=\phi=0 \label{8-8}
\end{equation}
\begin{equation}
\zeta=\frac{Q_E(r^2-y^2)}{\rho^4},
\>\>\>\>\>\>\>\> \eta=\frac{2Q_Ery}{\rho^4}\label{8-9}
\end{equation}
\begin{equation}
\zeta'=-\frac{2Q_Mry}{\rho^4}, \>\>\>\>\>\>\>\>  \eta'=\frac{Q_M(r^2-y^2)}{\rho^4}
\label{8-10}
\end{equation}
Eqs (\ref{8-7}) are identical with Eqs (\ref{B-33}), which means
that our metric becomes the metric of the Kerr-Newman solution
\cite{Ca1}. Also the $\zeta$ and $\eta$ of Eqs (\ref{B-34}) are
the sums $\zeta+\zeta'$ and $\eta+\eta'$, which are obtained from
Eqs (\ref{8-9}) and (\ref{8-10}).\vspace{3mm}

(3) Solution of Kerr \cite{Ke1}\vspace{3mm}

If
\begin{equation}
Q_E=Q_M=0
\label{8-11}
\end{equation}
Eq. (\ref{2-24}) gives $b=0$ and we get from Eqs (\ref{2-24}),
(\ref{2-25}) and (\ref{4-10})
\begin{equation}
 \rho^2=r^2+y^2,\>\>\>\>\>
\Delta=r^2-2Mr+y^2
\label{8-12}
\end{equation}
and from  Eqs (\ref{6-10}), (\ref{6-12})and (\ref{6-13})
\begin{equation}
 \xi=\phi=\zeta=\eta=\zeta'=\eta'=0
\label{8-13}
\end{equation}
Eqs (\ref{8-12}) and (\ref{B-37}) are identical, which means that
for  $Q_E=Q_M=0$ our solution becomes the solution of Kerr
\cite{Ke1}.\vspace{3mm}

(4) Metric and the axion and the dilaton field of the solution
given by Eqs (\ref{2-2}) and (\ref{2-3}) of this paper ( Class.
Quant. Grav. {\bf{23}}, 7591 (2006) Eqs (54)-(57) for
$\psi_{0}=0$, $AB=2Q_E^2$, $(\alpha-A)(\alpha-B)=2Q_M^2$,
$2M=\alpha-A-B$, $\psi=2\phi$ and $\alpha=b$ )\vspace{3mm}

If we introduce new parameters $M$, $Q_E$ and $Q_M$ by the relations
\begin{equation}
(b-A)(b-B)=2Q_M^2, \>\>\> AB=2Q_E^2, \>\>\> b-A-B=2M \label{8-14}
\end{equation}
the solution takes the form given in the Appendix namely its
metric is given by Eqs (\ref{2-18}) and (\ref{B-40}), its fields
axion and dilaton are given by Eq. (\ref{B-40'}) and its vector
field is given by Eqs (\ref{3-8})and (\ref{B-41}) We can show that
the metric, the axion field and the dilaton field of this solution
are obtained from the corresponding quantities of our solution for
\begin{equation}
a=0 \label{8-15}
\end{equation}
Indeed if we put $a=0$ in our solution we get from Eqs (\ref{2-24}), (\ref{2-25}),
(\ref{6-12}) and (\ref{6-13})
\begin{equation}
b=\frac{Q_E^2-Q_M^2}{M}, \>\>\>
\rho^2=r(r+b),\>\>\>
\Delta=r(r+b)-2Mr+2Q_M^2
\label{8-16}
\end{equation}
\begin{equation}
\xi=0, \>\>\>\>\> e^{-2\phi}=\frac{r+b}{r} \label{8-17}
\end{equation}
\begin{equation}
\zeta=\frac{Q_E}{(r+b)^2}, \>\>\> \eta=\zeta'=0, \>\>\>
\eta'=\frac{Q_M}{r(r+b)}\label{8-18}
\end{equation}
If we introduce new variables $r'$ and $b'$ by the relations
\begin{equation}
r=r'+b', \>\>\>\> b=-b' \label{8-19}
\end{equation}
Eqs (\ref{8-16})-(\ref{8-18}) become
\begin{equation}
b'=\frac{Q_M^2-Q_E^2}{M}, \>\>\>
\rho^2=r'(r'+b'),\>\>\>
\Delta=r'(r'+b')-2Mr'+2Q_E^2
\label{8-20}
\end{equation}
\begin{equation}
\xi=0, \>\>\>\>\> e^{2\phi}=\frac{r'+b'}{r'} \label{8-21}
\end{equation}
\begin{equation}
\zeta=\frac{Q_E}{r'^2}, \>\>\> \eta=\zeta'=0, \>\>\>
\eta'=\frac{Q_M}{r'(r'+b')}\label{8-22}
\end{equation}
If in (\ref{8-20})-(\ref{8-22}) we replace $r'$ by $r$ and $b'$ by
$b$ we see that Eqs (\ref{8-20}) and (\ref{B-40}) are identical
and the same think happens to Eqs Eqs (\ref{8-21}) and
(\ref{B-40'}). Therefore the metric and the axion and the dilaton
fields of the solution given by Eqs (\ref{2-2}) and (\ref{2-3})
\cite{Ky3} are obtained from the metric and the axion and the
dilaton fields of our solution for $a=0$. Also the solution given
by  Eqs (\ref{2-2}) and (\ref{2-3}) has a vector field with
electric charge $Q_E$ and magnetic charge $Q_M$, while our
solution has two vector fields one with electric charge $Q_E$ and
another with magnetic charge $Q_M$.

\vspace{3mm}

(5) The GM-GHS solution \cite{Gi1}\vspace{3mm}

If we put in our metric
\begin{equation}
a=Q_E=0 \label{8-23}
\end{equation}
we get from Eqs (\ref{2-24}) and (\ref{2-25})
\[b=-\frac{Q_M^2}{M}, \>\>\>\> \rho^2=r(r-\frac{Q_M^2}{M})\]
\begin{equation}
\Delta=r(r-\frac{Q_M^2}{M}) -2Mr+2Q_M^2=(r-2M)(r-\frac{Q_M^2}{M})
\label{8-24}
\end{equation}
and from  Eqs (\ref{6-12}) and (\ref{6-13})
\begin{equation}
\xi=0, \>\>\> e^{-2\phi}=1+\frac{b}{r}, \>\>\> \zeta=\eta=\zeta'=0,
\>\>\> \eta'= \frac{Q_M}{\rho^2} \label{8-25}
\end{equation}
From Eqs (\ref{8-24}), (\ref{8-25}), (\ref{B-50}) and (\ref{B-51})
we find that for $a=Q_E=0$ one of the vector fields of our
solution disappears and our solution becomes the GM-GHS solution
\cite{Gi1}.\vspace{3mm}

(6) Metric of the Reissner-Nordstr\"{o}m  solution for electric charge $Q_E$
and equal magnetic charge $Q_M$   \vspace{3mm}

If we put in our solution
\begin{equation}
a=0 \>\>\>\mbox{and} \>\>Q_E=Q_M \label{8-26}
\end{equation}
which implies according to Eq. (\ref{2-24}) that $b=0$, we get from Eqs (\ref{2-24})
and (\ref{2-25})
\begin{equation}
\rho^2=r^2, \>\>\>
\Delta=r^2 -2Mr+2Q_M^2=r^2 -2Mr+Q_E^2+Q_M^2
\label{8-27}
\end{equation}
and from Eqs (\ref{6-12}) and (\ref{6-13})
\begin{equation}
\xi=\phi=0, \>\>\>\> \zeta=\frac{Q_E}{r^2}, \>\>\>\> \eta=\zeta'=0,
 \>\>\>\>\eta'=\frac{Q_M}{r^2}
\label{8-28}
\end{equation}
 Eqs (\ref{8-27}) and (\ref{B-55}) are identical. Therefore the metric of the
 Reissner-Nordstr\"{o}m  solution for $Q_E=Q_M$ is obtained from the metric of
 our solution for $a=0$ and $Q_E=Q_M$. Also both solutions have electric charge $Q_E$
 and magnetic charge $Q_M$. However in the Reissner-Nordstr\"{o}m  solution the
 charges $Q_E$ and $Q_M$ belong to the same vector field, while in our case $Q_E$
 belongs to one vector field and $Q_M$ to another field.\vspace{3mm}

(7) The Schwarschild solution \vspace{3mm}

If we put in our solution
\begin{equation}
a=Q_E=Q_M=0 \label{8-29}
\end{equation}
from Eq. (\ref{2-24}) we get $b=0$ and Eqs (\ref{2-24}),
(\ref{2-25}), (\ref{6-12}) and (\ref{6-13}) give
\begin{equation}
\rho^2=r^2, \>\>\>\>\> \Delta=r^2 -2Mr
\label{8-30}
\end{equation}
\begin{equation}
\xi=\phi=\zeta=\eta=\zeta'=\eta'=0
\label{8-31}
\end{equation}
Eqs (\ref{8-30}) and (\ref{B-60}) are identical. Therefore if we
put $a=Q_E=Q_M=0$ in our solution we get the solution of
Schwarschild.

\section{``New Solution"}

It is well known that the equations of motion coming from the
action (\ref{3-1}) with $F'_{\mu\nu}=0$ are invariant under the
$SL(2,R)$ group of transformations \cite{Sh1}
\begin{equation}
\lambda \rightarrow\lambda'=\frac{\beta\lambda+\gamma}{\delta\lambda+\epsilon},
\>\>\>\>\>\>  \beta\epsilon-\gamma\delta=1
\label{9-1}
\end{equation}
\begin{equation}
F_+^{\mu\nu} \rightarrow -(\delta\lambda+\epsilon)F_{+}^{\mu\nu},
\>\>\>\>F_{-}^{\mu\nu} \rightarrow -(\delta\bar{\lambda}
+\epsilon)F_{-}^{\mu\nu} \label{9-2}
\end{equation}
This invariance known as $S$-duality also holds in the case we have an action of
the form (\ref{3-1}) with multiple vector fields provided that each vector field
transforms as above \cite{Ga1}. We get
\begin{equation}
\lambda'=\frac{\beta}{\delta}
-\frac{1}{\delta(\delta\lambda+\epsilon)} \label{9-3}
\end{equation}
The real constant ${\beta}{\delta}^{-1}$ makes a shift of the
axion field. Such a shift, which is allowed by the equations of
motion, will be omitted in which case we get for the $\lambda$ of
Eq (\ref{6-10})
\begin{equation}
\lambda'=
-\frac{1}{\delta(\delta\lambda+\epsilon)}=-\frac{r+iy}{\delta\{\epsilon
r-\delta y+i(\delta(r+b)+\epsilon y)\}}=\xi'+ie^{-2\phi'}
\label{9-4}
\end{equation}
where $\lambda'$ and $\phi'$ are the new axion and dilaton fields
respectively. But for this $\lambda'$ we get asymptotically
\begin{equation}
\lambda'\sim
-\frac{1}{\delta(\epsilon+i\delta)}=-\frac{\epsilon}{\delta(\delta^2+\epsilon^2)}
+\frac{i}{\delta^2+\epsilon^2} \label{9-5}
\end{equation}
which means that the asymptotic values of $\xi'$  and $\phi'$ are
not zero. To get a solution in which the axion and the dilaton
have zero asymptotic values we have according to Eq (\ref{6-16})
to make the replacements
\begin{equation}
\lambda'\rightarrow \lambda''=\{
\lambda'+\frac{\epsilon}{\delta(\delta^2+\epsilon^2)}\}(\delta^2+\epsilon^2)
 \label{9-6}
\end{equation}
\begin{equation}
-(\delta\lambda+\epsilon)F_{+}^{\mu\nu}\rightarrow f_{+}^{\mu\nu}
=-\frac{(\delta\lambda+\epsilon)}{\sqrt{\delta^2+\epsilon^2}}F_{+}^{\mu\nu},\>\>\>
-(\delta\bar{\lambda}+\epsilon)F_{-}^{\mu\nu}\rightarrow
f_{-}^{\mu\nu}
=-\frac{(\delta\bar{\lambda}+\epsilon)}{\sqrt{\delta^2+\epsilon^2}}F_{-}^{\mu\nu}
 \label{9-7}
\end{equation}
If we define $\delta'$ and $\epsilon'$ by the relations
\begin{equation}
\delta'=\frac{\delta}{\sqrt{\delta^2+\epsilon^2}}, \>\>\>\>
\epsilon'= \frac{\epsilon}{\sqrt{\delta^2+\epsilon^2}}\label{9-8}
\end{equation}
we get from Eqs (\ref{9-4}) and (\ref{9-6})
\begin{equation}
\lambda''=\frac{i\{(\epsilon'+i\delta')(r+iy)+\epsilon'b\}}{(\epsilon'+i\delta')(r+iy)
+i\delta'b}=\xi''+ie^{-2\phi''} \label{9-9}
\end{equation}
where
\begin{equation}
\xi''=\frac{b\{\delta'\epsilon'(2r+b)+(\epsilon'^2
-\delta'^2)y\}}{\delta'^2\{(r+b)^2+y^2\}+\epsilon'^2(r^2+y^2)+2\delta'\epsilon'by}
\label{9-10}
\end{equation}
\begin{equation}
e^{-2{\phi''}}=\frac{\rho^2}{\delta'^2\{(r+b)^2+y^2\}+\epsilon'^2(r^2+y^2)
+2\delta'\epsilon'by}
\label{9-11}
\end{equation}
Taking into account Eqs (\ref{9-8}) Eqs (\ref{9-7}) give for the
$f_{\mu\nu}$ field and the second vector field $f'_{\mu\nu}$
\cite{Se2}-\cite{Sh1}
\begin{equation}
f_{\mu\nu}=-(\delta'\xi+\epsilon')F_{\mu\nu} +\delta'
e^{-2\phi}F^\ast_{\mu\nu} \label{9-12}
\end{equation}
\begin{equation}
f'_{\mu\nu}=-(\delta'\xi+\epsilon')F'_{\mu\nu} +\delta'
e^{-2\phi}F'^\ast_{\mu\nu} \label{9-13}
\end{equation}
From the $F_{\mu\nu}$ field of Eq (\ref{3-8}) we can calculate the
${F^\ast}_{\mu\nu}$ field using Eq. (\ref{3-2}). We get
\[{F^\ast}_{rt}=\eta,\>\>\> {F^\ast}_{r\phi}=-a\sin^2\theta\eta, \>\>\>
{F^\ast}_{\theta t}=a\sin\theta\zeta, \>\>\>
{F^\ast}_{\theta\phi}=-\{r(r+b)+a^2\}\sin\theta\zeta\]
\begin{equation}
{F^\ast}_{r\theta}={F^\ast}_{t\phi}=0  \label{9-14}
\end{equation}
which are obtained from the $F_{\mu\nu}$ field if we make the
substitutions $\zeta\rightarrow\eta$ and $\eta\rightarrow-\zeta$.
Generally if $F_{\mu\nu}$ is of the form of Eq. (\ref{3-8}) and
$g_{\mu\nu}$ is given by expression (\ref{2-18}) with $\rho^2$
given by Eq. (\ref{2-24}) its dual
$F^\ast_{\mu\nu}=\frac{g_{\mu\sigma}
g_{n\tau}}{2\sqrt{-g}}\epsilon^{\sigma\tau \chi \psi}F_{\chi\psi}$
is obtained from $F_{\mu\nu}$ by the substitution
\begin{equation}
 \zeta \rightarrow \eta, \>\>\>\>\> \eta \rightarrow -\zeta\label{9-14'}
\end{equation}
Also from Eq. (\ref{3-9}) we get for the second vector field
\[{F'^\ast}_{rt}=\eta',\>\>\> {F'^\ast}_{r\phi}=-a\sin^2\theta\eta', \>\>\>
{F'^\ast}_{\theta t}=a\sin\theta\zeta', \>\>\>
{F'^\ast}_{\theta\phi}=-\{r(r+b)+a^2\}\sin\theta\zeta'\]
\begin{equation}
{F'^\ast}_{r\theta}={F'^\ast}_{t\phi}=0  \label{9-15}
\end{equation}
The vector fields $f_{\mu\nu}$ and $f'_{\mu\nu}$ of the new
solution are obtained from Eqs (\ref{3-8}), (\ref{3-9}),
(\ref{6-12}), (\ref{6-13}) and (\ref{9-12})-(\ref{9-15}). We get
\[f_{rt}=\sigma,\>\>\> f_{r\phi}=-a\sin^2\theta\sigma, \>\>\>
f_{\theta t}=-a\sin\theta\tau, \>\>\>
f_{\theta\phi}=\{r(r+b)+a^2\}\sin\theta\tau\]
\begin{equation}
f_{r\theta}=f_{t\phi}=0  \label{9-16}
\end{equation}
\[f'_{rt}=\sigma',\>\>\> f'_{r\phi}=-a\sin^2\theta\sigma', \>\>\>
f'_{\theta t}=-a\sin\theta\tau', \>\>\>
f'_{\theta\phi}=\{r(r+b)+a^2\}\sin\theta\tau'\]
\begin{equation}
f'_{r\theta}=f'_{t\phi}=0  \label{9-17}
\end{equation}
where
\begin{equation}
\sigma=-\frac{\{\delta'by+\epsilon'(r^2+y^2)\}\zeta-\delta'\rho^2\eta}{r^2+y^2}
=\frac{Q_E}{\rho^4}\{\delta'y(2r+b)-\epsilon'(r^2-y^2)\}
\label{9-18}
\end{equation}
\begin{equation}
\tau=-\frac{\{\delta'by+\epsilon'(r^2+y^2)\}\eta+\delta'\rho^2\zeta}{r^2+y^2}=
-\frac{Q_E}{\rho^4}\{\delta'\{r(r+b)-y^2\}+2\epsilon'ry\}
 \label{9-19}
\end{equation}
\begin{equation}
\sigma'=-\frac{\{\delta'by+\epsilon'(r^2+y^2)\}\zeta'-\delta'\rho^2\eta'}{r^2+y^2}
=\frac{Q_M}{\rho^4}\{\delta'\{(r+b)^2-y^2\}+\epsilon'(2r+b)y\}
\label{9-20}
\end{equation}
\begin{equation}
\tau'=-\frac{\{\delta'by+\epsilon'(r^2+y^2)\}\eta'+\delta'\rho^2\zeta'}{r^2+y^2}=
\frac{Q_M}{\rho^4}\{2\delta'y(r+b)-\epsilon'\{r(r+b)-y^2\}\}
\label{9-21}
\end{equation}

For large $r$ we have
\begin{equation}
f_{rt}\sim-\frac{Q_E\epsilon'}{r^2}+O(\frac{1}{r^3}), \>\>\>\>
f_{\theta\phi}\sim-Q_E\delta'\sin\theta+O(\frac{1}{r})
\label{9-22}
\end{equation}
\begin{equation}
f'_{rt}\sim\frac{Q_M\delta'}{r^2}+O(\frac{1}{r^3}), \>\>\>\>
f'_{\theta\phi}\sim-Q_M\epsilon'\sin\theta+O(\frac{1}{r})
\label{9-23}
\end{equation}
Therefore the electric charge $q_E$ and the magnetic charge $q_M$
of the first vector field are
\begin{equation}
q_E=-Q_E\epsilon', \>\>\>\>\> q_M=-Q_E\delta'
\label{9-24}
\end{equation}
and the electric charge $q'_E$ and the magnetic charge $q'_M$ of
the second vector field are
\begin{equation}
q'_E=Q_M\delta', \>\>\>\>\> q'_M=-Q_M\epsilon' \label{9-25}
\end{equation}
The following relations are satisfied
\begin{equation}
q_Eq'_E+q_Mq'_M=0, \>\>\>\> q_E^2+q_M^2=Q_E^2, \>\>\>\>
{q'_E}^2+{q'_M}^2=Q_M^2 \label{9-26}
\end{equation}
where $Q_E$ and $Q_M$ are arbitrary.  Therefore only three of the
charges $q_E$, $q_M$, $q'_E$ and $q'_M$ can take arbitrary values.
Using Eqs (\ref{9-24})and (\ref{9-25}) we can express $\sigma$,
$\tau$, $\sigma'$ and $\tau'$ in terms of the charges $q_E$,
$q_M$, $q'_E$ and $q'_M$. We get
\begin{equation}
\sigma=\frac{q_E(r^2-y^2)-q_My(2r+b)}{\rho^4}, \>\>\>\>
\tau=\frac{q_M\{r(r+b)-y^2\}+2q_Ery}{\rho^4} \label{9-27}
\end{equation}
\begin{equation}
\sigma'=\frac{q'_E\{(r+b)^2-y^2)\}-q'_My(2r+b)}{\rho^4}, \>\>\>\>
\tau'=\frac{q'_M\{r(r+b)-y^2\}+2q'_Ey(r+b)}{\rho^4} \label{9-28}
\end{equation}

The vector fields $f_{\mu\nu}$ and $f'_{\mu\nu}$ can be obtained
from the vector potentials $a_\mu$ and $a'_\mu$ respectively where
\begin{equation}
a_t= \frac{Q_E}{\rho^2}(\epsilon'r-\delta'y), \>\>\>
a_\phi=\frac{Q_E}{\rho^2}\{-\epsilon'ra\sin^2\theta+\delta'\{r(r+b)
+a^2\}\cos\theta\}, \>\>\> a_r=a_\theta=0 \label{9-29}
\end{equation}
\[a'_t= -\frac{Q_M}{\rho^2}\{\epsilon'a\cos\theta+\delta'(r+b)\},\>\>\>
a'_\phi=\frac{Q_M}{\rho^2}\{\epsilon'\{r(r+b)+a^2\}\cos\theta\]
\begin{equation}
+\delta'(r+b)a\sin^2\theta\}, \>\>\> a'_r=a'_\theta=0
\label{9-30}
\end{equation}

For the ``new solution" equations of the type (\ref{5-6}) must
hold. If we write for the first vector field
\begin{equation}
e^{-2\phi''}\sigma+\xi''\tau=\partial_yS, \>\>\>\>
e^{-2\phi''}\tau-\xi''\sigma=-\partial_rS, \>\>\>\> \label{9-31}
\end{equation}
and for the second vector field
\begin{equation}
e^{-2\phi''}\sigma'+\xi''\tau'=\partial_yS', \>\>\>\>
e^{-2\phi''}\tau'-\xi''\sigma'=-\partial_rS' \label{9-32}
\end{equation}
and use  Eqs (\ref{9-10}), (\ref{9-11}) and
(\ref{9-18})-(\ref{9-21}) we get
\begin{equation}
S=-\frac{Q_E}{\rho^2}(\delta'r+\epsilon'y) \label{9-32}
\end{equation}
\begin{equation}
S'=\frac{Q_M}{\rho^2}\{\delta'y-\epsilon'(r+b)\}\label{9-32'}
\end{equation}
which are of the form of Eqs (\ref{5-8}) and (\ref{5-10})
respectively.

Eq. (\ref{9-9}) can be written in the form
\begin{equation}
\lambda''=\frac{i\{r+iy+(1-d)b\}}{r+iy+db} \label{9-33}
\end{equation}
where
\begin{equation}
d=\frac{i\delta'}{\epsilon'+i\delta'} \label{9-34}
\end{equation}
The quantity $\lambda$ of our first solution given by Eq.
(\ref{6-10}) is again of the form of Eq. (\ref{9-33}) with
\begin{equation}
d=0 \label{9-35}
\end{equation}
All solutions with $\phi\neq 0$ which are mentioned in this paper
have a $\lambda$ of the form of Eq. (\ref{9-33}) with $d$ a real
or a complex constant,. In the Appendix we calculate the constant
$d$ for all solutions with $\phi\neq0$ which are listed there. For
the fields $\xi$ and $\phi$ coming from an expression of the form
of Eq. (\ref{9-33}) we get $\xi_\infty=\phi_\infty=0$, where
$\xi_\infty$ and $\phi_\infty$ are their asymptotic values
respectively. Therefore before we check if the $\lambda$ of a
solution can be written in the form of Eq. (\ref{9-33}) we have to
take out the asymptotic values  $\xi_\infty$ and $\phi_\infty$, if
they are non-vanishing.

\section{Special cases II}

If in  Eqs (\ref{9-9}) and (\ref{9-18})-(\ref{9-21}) we put
\begin{equation}
\delta'=0, \>\>\>\>\>\> \epsilon'=-1 \label{10-1}
\end{equation}
we get expressions (\ref{6-10}) and (\ref{6-13}). Therefore our
solution is a special case of our ``new solution" and it is
obtained for $\delta'=0$ and $\epsilon'=-1$. This means that  all
solutions of Section 8, which are special cases of our solution,
are also special cases of our ``new solution". In addition we
shall show that two more solution \cite{Se2},\cite{Sh1} are
obtained from our ``new solution" for certain values of its
parameters. \vspace{3mm}

(1) Solution of Sen with both charges electric and magnetic
\cite{Se2} \vspace{3mm}

If in our ``new solution" we put
\begin{equation}
Q_M=0 \label{10-2}
\end{equation}
we get from Eqs (\ref{9-17}), (\ref{9-20}) and (\ref{9-21})
\begin{equation}
f'_{\mu\nu}=0 \label{10-3}
\end{equation}
and from Eqs (\ref{2-24}) and (\ref{2-25})
\begin{equation}
\rho^2=r(r+\frac{Q_E^2}{M})+y^2,\>\>\>
\Delta=r(r+\frac{Q_E^2}{M})-2Mr+y^2 \label{10-4}
\end{equation}
Also if in Eqs (\ref{10-4}), (\ref{9-9})-(\ref{9-11}),
(\ref{9-18}) and (\ref{9-19}) we put
\begin{equation}
Q_E^2=q_E^2+q_M^2, \>\>\>\>
\delta'=-\frac{q_M}{\sqrt{q_E^2+q_M^2}}, \>\>\>\>
\epsilon'=-\frac{q_E}{\sqrt{q_E^2+q_M^2}} \label{10-5}
\end{equation}
and in the resulting expressions we replace $q_E$ by $Q_E$ and
$q_M$ by $Q_M$ we get Eqs (\ref{B-26})-(\ref{B-30}). Therefore the
solution of Sen with both charges electric and magnetic \cite{Se2}
is obtained from our ``new solution" for $Q_M=0$. \vspace{3mm}

(2) Solution of Shapere, Trivedi and Wilczek
\cite{Sh1}\vspace{3mm}

If in our ``new solution" we put
\begin{equation}
a=Q_E=0 \label{10-6}
\end{equation}
we get from Eqs (\ref{2-24}) and (\ref{2-25})
\begin{equation}
b=-\frac{Q_M^2}{M}, \>\>\>\>\> \rho^2=r(r-\frac{Q_M^2}{M})
\label{10-7}
\end{equation}
\begin{equation}
\Delta=r(r-\frac{Q_M^2 }{M})-2Mr+2Q_M^2=(r-2M)(r-\frac{Q_M^2 }{M})
 \label{10-7'}
\end{equation}
 from Eqs (\ref{9-16}), (\ref{9-18}) and (\ref{9-19})
\begin{equation}
f_{\mu\nu}=0 \label{10-8}
\end{equation}
from Eq (\ref{9-9})
\begin{equation}
\lambda''=\frac{i\{(\epsilon'+i\delta')r+\epsilon'b\}}{(\epsilon'+i\delta')r
+i\delta'b} \label{10-9}
\end{equation}
and from Eqs (\ref{9-20}) and (\ref{9-21})
\begin{equation}
\sigma'= \frac{Q_M\delta'}{r^2}, \>\>\>\>\>
\tau'=-\frac{Q_M\epsilon'}{\rho^2} \label{10-10}
\end{equation}
If we define $Q'_E$ and $Q'_M$ by the relations
\begin{equation}
Q'_E=Q_M\delta', \>\>\>\>\> Q'_M=-Q_M\epsilon' \label{10-11}
\end{equation}
which imply according to Eq (\ref{9-8}) that
\begin{equation}
{Q'_E}^2+{Q'_M}^2=Q_M^2 \label{10-12}
\end{equation}
using these relations eliminate $\delta'$, $\epsilon'$ and $Q_M$
from the expressions (\ref{10-7}), (\ref{10-7'}), (\ref{10-9}) and
(\ref{10-10}), write $\zeta$ instead of $\sigma'$ and $\eta$
instead of $\tau'$ and subsequently drop the primes
 we get the
relations
\begin{equation}
\rho^2=r(r-\frac{Q_E^2+Q_M^2}{M}) \label{10-13}
\end{equation}
\begin{equation}
\Delta=r(r-\frac{Q_E^2+Q_M^2
}{M})-2Mr+2(Q_E^2+Q_M^2)=(r-2M)(r-\frac{Q_E^2+Q_M^2 }{M})
\label{10-13'}
\end{equation}

\begin{equation}
\lambda=\frac{i\{(Q_M-iQ_E)r+Q_M b\}}{(Q_M-iQ_E)r-iQ_E
b}\label{10-14}
\end{equation}

\begin{equation}
\zeta=\frac{Q_E}{r^2},\>\>\>\>\eta=\frac{Q_M}{\rho^2}\label{10-15}
\end{equation}
Eqs (\ref{10-13})-(\ref{10-15}) are identical with Eqs
(\ref{B-44})-(\ref{B-45}) and (\ref{B-46}). Therefore the STW
solution \cite{Sh1} is obtained from our ``new solution" for
$a=Q_E=0$.\vspace{1cm}

\section{Energy Conditions}

Using Einstein's equations and the
expressions(\ref{4-1})-(\ref{4-4}) and (\ref{7-14}) for the
components $R_{\mu\nu}$ of the Ricci tensor and the Ricci scalar
$R$ we find the components $T_{\mu\nu}$ of the energy-momentum
tensor
\begin{equation}
T_{\mu\nu}=R_{\mu\nu}-\frac{g_{\mu\nu}}{2}R \label{7'-1}
\end{equation}
The eingevalues $w_\mu$ of the tensor $T_\mu^\nu$ are obtained if
we equate to zero the determinant of the eingevalue equation
\begin{equation}
(T_\mu^\nu-w\delta_\mu^\nu)u^\mu=0 \label{7'-2}
\end{equation}
namely from the relation
 \[ |T_\mu^\nu-w\delta_\mu^\nu|=0=\]
\begin{equation} \left|
\begin{array}{lccr}
R_{tt}g^{tt}+R_{t\phi}g^{\phi t}-\frac{R}{2}-w
&0{\hspace{-10mm}}&0&R_{tt}g^{t\phi}+R_{t\phi}g^{\phi\phi}\\
0&R_{rr}g^{rr}-\frac{R}{2}-w{\hspace{-10mm}}&0&0\\
0&0{\hspace{-10mm}}&R_{\theta\theta}g^{\theta\theta}-\frac{R}{2}
-w&0\\
R_{\phi\phi}g^{\phi t}+R_{\phi t}g^{tt}&0{\hspace{-10mm}}&0
&R_{\phi\phi}g^{\phi\phi}+R_{\phi t}g^{t\phi}-\frac{R}{2}-w
\end{array} \right|
\label{7'-3}
\end{equation}
Using the relations
\begin{equation}
R_{tt}g^{tt}+R_{\phi\phi}g^{\phi\phi}+2R_{t\phi}g^{t\phi}=0
\label{7'-4}
\end{equation}
\begin{equation}
R_{tt}R_{\phi\phi}-(R_{t\phi})^2=
\frac{K^2\sin^2\theta(\Delta+a^2\sin^2\theta)}{\rho^{12}}\label{7'-5}
\end{equation}
\begin{equation}
g^{tt}g^{\phi\phi}-(g^{t\phi})^2=-\frac{1}{(\Delta+a^2\sin^2\theta)\sin^2\theta}
 \label{7'-6}
\end{equation}
which hold for any metric of the form of Eq. (\ref{2-18}), we find
that the energy density $\mu=-w_0$, where $w_o$ is the eingevalue
which corresponds to the time-like eigenvector  and the principal
pressures $p_i=w_i\>\>\>i=1,\> 2,\> 3$ of our solution are
\begin{equation}
\mu=\frac{K}{\rho^6}+\frac{R}{2}
 \label{7'-7}
\end{equation}
\begin{equation}
p_1=\frac{K}{\rho^6}-\frac{R}{2}, \>\>\>
p_2=\frac{K}{\rho^6}-\frac{R}{2}+\frac{a^2b^2\sin^2\theta}{2\rho^6},
\>\>\>
p_3=-\frac{K}{\rho^6}+\frac{R}{2}-\frac{a^2b^2\sin^2\theta}{2\rho^6}
 \label{7'-8}
\end{equation}
with $K$  given by Eq (\ref{4-4}).

For a metric of the form of Eqs (\ref{2-18}) and (\ref{2-19}) the
event horizons are the surfaces given by  Eq (\ref{7-12}),   from
which we get for the outer horizon $r_+^H$
\begin{equation}
2r_+^H+A+B=\sqrt{(A-B)^2-4a^2} \label{7'-9}
\end{equation}
The Ricci scalar $R$ given by Eq (\ref{7-14}) becomes on the outer
horizon
\begin{equation}
R=\frac{b^2(\Delta+2a^2b^2\sin^2\theta)}{2\rho^2}
=\frac{a^2b^2\sin^2\theta}{2\rho^2}\geq0 \label{7'-10}
\end{equation}
Also outside the outer horizon where $r=r_+^H+r'$, with $r'>0$, we
have
\begin{equation}
\Delta+a^2\sin^2\theta=r'(r'+\sqrt{(A-B)^2-4a^2})>0
 \label{7'-11}
\end{equation}
Therefore for all solutions with metric of the form given by Eqs
(\ref{2-18}) and (\ref{2-19}) ( obviously Eq. (\ref{2-19}) can be
replaced by Eqs (\ref{2-23})-(\ref{2-25})) we have outside and on
the outer horizon
\begin{equation}
\Delta+a^2\sin^2\theta\geq 0 \>\>\>\>\> \mbox{and} \>\>\>\>\>
R\geq 0
 \label{7'-12}
\end{equation}
For our solution we have
\begin{equation}
b>0, \>\>\>\>\> A<0, \>\>\>\>\> B<0
 \label{7'-13}
\end{equation}
so that we get from Eq (\ref{2-23}) the relations
\begin{equation}
2Q_E^2=(b-A)(b-B)>0, \>\>\>\>\> 2Q_M^2=AB>0
 \label{7'-14}
\end{equation}
which give according to Eq. (\ref{4-4}) the relation
\begin{equation}
K>0
 \label{7'-15}
\end{equation}
everywhere. Therefore we get from Eqs (\ref{7'-7}), (\ref{7'-8}),
(\ref{7'-12}) and (\ref{7'-15}) outside and on the outer horizon
\[\mu=\frac{K}{\rho^6}+\frac{R}{2}>0, \>\>\>\>
\mu+p_1=\frac{2K}{\rho^6}>0, \>\>\>\>
\mu+p_2=\frac{2K}{\rho^6}+\frac{a^2b^2\sin^2\theta}{2\rho^6}>0\]
\[\mu+p_3=R-\frac{a^2b^2\sin^2\theta}{2\rho^6}
=\frac{b^2(\Delta+a^2\sin^2\theta)}{2\rho^6}\geq0, \>\>\>\>
\mu-p_1=R\geq0\]
\begin{equation}
\mu-p_2=R-\frac{a^2b^2\sin^2\theta}{2\rho^6}
=\frac{b^2(\Delta+a^2\sin^2\theta)}{2\rho^6}\geq0, \>\>\>\>
\mu-p_3=\frac{2K}{\rho^6}+\frac{a^2b^2\sin^2\theta}{2\rho^6}>0
\label{7'-16}
\end{equation}
and the dominant energy conditions, which are defined by the
relations \cite{Ha1}
\begin{equation}
\mu\geq0, \>\>\>\>\>\> -\mu \leq p_i \leq \mu, \>\>\> i=1,\> 2, \>
3\label{7'-17}
\end{equation}
are satisfied. Of coarse the weak energy conditions, which are
defined by the relations \cite{Ha1}
\begin{equation}
\mu\geq0, \>\>\>\>\>\> \mu +p_1\geq 0, \>\>\> i=1,\> 2, \>
3\label{7'-18}
\end{equation}
are satisfied. Also we have
\begin{equation}
\mu+p_1+p_2+p_3=\frac{2K}{\rho^6}>0  \label{7'-19}
\end{equation}
and the strong energy conditions, which are defined by the
relations\cite{Ha1}
\begin{equation}
\mu +p_1\geq 0, \>\>\> i=1,\> 2,\> 3, \>\>\>\> \mu+p_1+p_2+p_3
\geq 0 \label{7'-20}
\end{equation}
are satisfied. Therefore our solution and also our ``new
solution", whose metric is obtained from the metric of our
solution if we make the substitutions
\begin{equation}
Q_E^2 \longrightarrow q_E^2 + q_M^2 \equiv Q^2, \>\>\>\>\>\> Q_M^2
\longrightarrow {q'}_E^2 + {q'}_M^2 \equiv {Q'}^2 \label{7'-20'}
\end{equation}
satisfy the dominant, the weak and the strong energy conditions
outside and on the outer horizon.

We observe that all solutions with metric of the form of Eq.
(\ref{2-18}) with $\rho^2$ and $\Delta$ given by Eq. (\ref{2-19})
or by Eqs (\ref{2-23})-(\ref{2-25}) satisfy all energy conditions
outside and on the outer horizon if the relations
$\Delta+a^2\sin^2\theta\geq 0$, $R\geq 0$ and $K\geq 0$ are
satisfied outside and on the outer horizon, where $R$ is the Ricci
scalar of the metric and $K$ is defined by Eq. (\ref{4-4}). But we
proved before that all solutions with metric of the above form
satisfy the relation $\Delta+a^2\sin^2\theta\geq 0$ and $R\geq 0$
outside and on the outer horizon. Also we observe that if we
define $Q_E$, $Q_M$ and $K$ by the relations (\ref{2-23}) and
(\ref{4-4}) and take
\begin{equation}
A< 0,\>\>\> \mbox{and} \>\>\> B\leq 0, \>\>\> b\geq 0\>\>\>
\mbox{or} \>\>\> b=B<0 \label{7'-21}
\end{equation}
we get $K\geq 0$ everywhere. Therefore all solutions with metric
of the form of Eq. (\ref{2-18}) with $\rho^2$ and $\Delta$ given
by Eq. (\ref{2-19}) or by Eqs (\ref{2-23})-(\ref{2-25}), whose
parameters $A$, $B$ and $b$ satisfy the relations (\ref{7'-21}),
satisfy the dominant, the weak and the strong energy conditions
outside and on the outer horizon. This happens to all solutions
given in the Appendix except the Kerr-Newman solution and the
 Reissner-Nordstr\"{o}m solution if their electric charge $Q_E$
and magnetic charge $Q_M$ satisfy the relation $Q_E\neq Q_M$, for
which we cannot reach this conclusion because Eqs
(\ref{2-23})-(\ref{2-25}) are not satisfied. But we can easily
show that these solutions also satisfied all energy conditions. To
do that consider a Kerr-Newman solution with charges $Q'_E$ and
$Q'_M$ for which $Q'_E \neq Q'_M$ and our solution with charges
$Q_E$ and $Q_M$ which satisfy the relations
\begin{equation}
Q_E^2=Q_M^2=\frac{{Q'}_E^2+{Q'}_M^2}{2}\label{7'-22}
\end{equation}
Both solutions have the same metric, which means that they have
the same components of the Ricci tensor $R_{\mu\nu}$ and the same
Ricci scalar $R$. Therefore if in $R_{\mu\nu}$ and $R$ of Eqs
(\ref{4-1})-(\ref{4-4}) and (\ref{7-14}) we make the substitutions
of  Eq. (\ref{7'-22}) we get the $R_{\mu\nu}$ and $R$ of the
Kerr-Newman solution with $Q'_E \neq Q'_M$. More specifically we
get since $b=0$
\begin{equation}
 K=({Q'_E}^2+{Q'_M}^2)(r^2+a^2\cos^2\theta) \>\>\>\> \mbox{and}
\>\>\>\> R=0\label{7'-23}
\end{equation}
Therefore we get from  Eqs (\ref{7'-7}) and (\ref{7'-8})
\begin{equation}
 \mu=p_1=p_2=-p_3=\frac{K}{\rho^6}\label{7'-24}
\end{equation}
Since we have
\[ \mu>0, \>\>
\mu+p_1=\mu+p_2=\mu-p_3=\mu+p_1+p_2+p_3=\frac{2K}{\rho^6}>0\]
\begin{equation}
 \mu+p_3=\mu-p_1=\mu-p_2=0 \label{7'-24}
\end{equation}
 Eqs (\ref{7'-17}), (\ref{7'-18}) and (\ref{7'-20}) are satisfied everywhere,
which means that the Kerr-Newman solution satisfies all energy
conditions.

Also since the Reissner-Nordstr\"{o}m solution is obtained from
the Kerr-Newman solution for $a=0$ for which
$K=({Q'_E}^2+{Q'_M}^2)r^2>0$ Eqs (\ref{7'-17}), (\ref{7'-18}) and
(\ref{7'-20}) are satisfied everywhere, which means that the
Reissner-Nordstr\"{o}m solution for arbitrary $Q_E$ and $Q_M$
satisfies all energy conditions.

\section{Mass Formulae}

In this section we shall consider mass formulae of solutions with
metric $g_{\mu\nu}$ of the form of Eqs (\ref{2-18}), (\ref{2-24})
and (\ref{2-25}) and of the form of Eqs (\ref{2-18}) and
(\ref{2-25'}) and angular momentum $J=Ma$, where $a$ is a non-zero
or a zero constant. To find the expressions for $M$ and $dM$
assuming that we have a metric of the form of Eqs (\ref{2-18}),
(\ref{2-24}) and (\ref{2-25}) and angular momentum $J\neq0$ we
prossed as follows: We get from Eq (\ref{7-13})
\begin{equation}
r_+^H r_-^H -a^2=2Q_M^2 \label{11-1}
\end{equation}
from Eqs (\ref{7-8}) and (\ref{7-13}) for $a\neq0$
\begin{equation}
a(r_+^H +r_-^H +b)=2Ma=2J \label{11-2}
\end{equation}
and from Eq (\ref{7-18})
\begin{equation}
r_+^H(r_+^H +b)=\frac{N}{4\pi} -a^2 \label{11-3}
\end{equation}
The above system is equivalent to the system formed by Eq
(\ref{11-1}) and the relations
\begin{equation}
\frac{2Jr_+^H}{a}=\frac{N}{4\pi} +2Q_M^2 \label{11-4}
\end{equation}
\begin{equation}
\frac{2Jr_-^H}{a}=\frac{4J^2}{a^2}-\frac{N}{4\pi} -2Q_M^2
-\frac{2Jb}{a}= \frac{4J^2}{a^2}-\frac{N}{4\pi}
-2Q_E^2\label{11-5}
\end{equation}
In writing Eq (\ref{11-5}) we used the relations $J=Ma$ and
$b=\frac{Q_E^2 -Q_M^2}{M}$. Multiplying Eqs (\ref{11-4}) and
(\ref{11-5}) by parts and using Eq (\ref{11-1}) and the relation
$J=Ma$ we get the mass formula
\begin{equation}
M=\{ \frac {4\pi J^2}{N}+\frac{\pi}{N}(\frac{N}{4\pi}
+2Q_E^2)(\frac{N}{4\pi} +2Q_M^2 )\}^{\frac{1}{2}} \label{11-6}
\end{equation}

The above $M$ is a homogeneous function of degree $\frac{1}{2}$ of
the variables $x^1=J$, $x^2=N$, $x^3=Q_E^2$ and $x^4=Q_M^2$.
Therefore by applying to it Euler's theorem on homogeneous
functions and by taking the differential of $M$ we get
respectively
\begin{equation}
 M=2x^i \frac{\partial M}{\partial x^i} \>\>\>\> \mbox{and} \>\>\>\>
dM=\frac{\partial M}{\partial x^i}dx^i \label{11-7}
\end{equation}
 Since we have
\begin{equation}
\frac{\partial M}{\partial J}=\frac{4\pi
J}{MN}=\frac{a}{r_+^H(r_+^H +b)+a^2}\label{11-8}
\end{equation}
\begin{equation}
 \frac{\partial
M}{\partial Q_E^2}=\frac{1}{4M}(1+\frac{8\pi
Q_M^2}{N})=\frac{r_+^H}{2\{r_+^H(r_+^H +b)+a^2\}} \label{11-9}
\end{equation}
\begin{equation}
\frac{\partial M}{\partial Q_M^2}=\frac{1}{4M}(1+\frac{8\pi
{Q_E}^2}{N})=\frac{ r_+^H+b}{2\{r_+^H(r_+^H +b)+a^2\}}
\label{11-10}
\end{equation}
\begin{equation}
\frac{\partial M}{\partial
N}=\frac{1}{2M}(\frac{1}{16\pi}-\frac{4\pi J^2}{N^2}-\frac{4\pi
Q_E^2 Q_M^2}{N^2})=\frac{r_+^H -r_-^H}{16\pi\{r_+^H(r_+^H
+b)+a^2\}}\label{11-11}
\end{equation}
we find that $M$ and $dM$ are given by the relations
\begin{equation}
M=2J \Omega+Q_E \phi_E +Q_M \phi_M+\frac{\kappa}{4\pi}N
\label{11-12}
\end{equation}
\begin{equation}
dM=\Omega dJ+\phi_E dQ_E+\phi_M dQ_M+\frac{\kappa}{8\pi}dN
\label{11-13}
\end{equation}
where the area of the outer horizon $N$, the angular velocity
$\Omega$ and the surface gravity $\kappa$ are given by Eqs
(\ref{7-18}), (\ref{7-22}) and (\ref{7-20}) respectively, and
$\phi_E$ and $\phi_M$, which are given by the relations
\begin{equation}
\phi_E=\frac{Q_Er_+^H}{r_+^H(r_+^H +b)+a^2}, \>\>\>\>
\phi_M=\frac{Q_M(r_+^h +b)}{r_+^H(r_+^H +b)+a^2} \label{11-14}
\end{equation}
are the electric potential and the magnetic potential on the outer
horizon respectively. Eq. (\ref{11-12}) is a mass formula of
Smarr's type \cite{Sm1} and Eq. (\ref{11-13}) a differential mass
formula. Expressions  (\ref{11-6}), (\ref{11-12}) and
(\ref{11-13}) are the mass formulae of our solution.

According to Eq. (\ref{9-26}) the metric $g_{\mu\nu}$ of our ``new
solution" is obtained from the metric of our solution if we make
the substitutions of Eq. (\ref{7'-20'}). Also its angular momentum
is again $J=Ma$. Since the mass formulae were  obtained from the
metric and the relation $J=Ma$ we get for our ``new solution" the
relations
\begin{equation}
 M=\{\frac{4\pi J^2}{N}+\frac{\pi}{N}(\frac{N}{4\pi}
+2Q^2)(\frac{N}{4\pi} +2{Q'}^2 )\}^{\frac{1}{2}}\label{11-16}
\end{equation}
\begin{equation}
M=2J \Omega+Q \phi +Q' \phi'+\frac{\kappa}{4\pi}N \label{11-17}
\end{equation}
\begin{equation}
 dM=\Omega dJ+\phi dQ+\phi' dQ'+\frac{\kappa}{8\pi}dN \label{11-18}
\end{equation}
where the area of the outer horizon $N$ the angular velocity
$\Omega$ and the surface gravity $\kappa$ are given by Eqs
(\ref{7-18}), (\ref{7-22}) and (\ref{7-20}) respectively, the
potentials $\phi$ and $\phi'$ by the relations
\begin{equation}
\phi=\frac{Qr_+^H}{r_+^H(r_+^H +b)+a^2} \>\>\>\> \mbox{and}
\>\>\>\> \phi'=\frac{Q'(r_+^h +b)}{r_+^H(r_+^H +b)+a^2}
\label{11-19}
\end{equation}
and $b$ by the relation
\begin{equation}
b=\frac{Q^2-{Q'}^2}{M}\label{11-20}
\end{equation}

The solution of Sen with electric charge only is obtained from our
solution for $Q_M=0$. Therefore the mass formulae of this solution
of Sen are obtained from the mass formulae of our solution for
$Q_M=0$ that is from the expressions
\begin{equation}
M=\{\frac{4\pi
J^2}{N}+\frac{1}{4}(\frac{N}{4\pi}+2Q_E^2)\}^{\frac{1}{2}}\label{11-21}
\end{equation}
\begin{equation}
M=2J\Omega +Q_E \phi_E+\frac{\kappa}{4\pi}N \label{11-22}
\end{equation}
\begin{equation}
dM=\Omega dJ+\phi_E dQ_E+\frac{\kappa}{8\pi}dN \label{11-23}
\end{equation}
which are obtained from Eqs (\ref{11-6}), (\ref{11-12}) and
(\ref{11-13}) respectively for $Q_M=0$. The quantities $J$,
$\Omega$, $N$, $\kappa$ and $\phi_E$ in Eqs
(\ref{11-21})-(\ref{11-23}) are given by Eqs (\ref{7-8}),
(\ref{7-22}), (\ref{7-18}) (\ref{7-20}) and (\ref{11-14})
respectively and $b=\frac{Q_E^2}{M}$.

Since the solution of Sen with electric charge $q_E$ and magnetic
charge $q_M$ is obtained from our ``new solution" for $Q'=0$ the
mass formulae of this solution of Sen are obtained if we make the
substitutions $Q'=0$ and $Q^2=q_E^2+q_M^2$ in the mass formulae of
our ``new solution", which are given by Eqs
(\ref{11-16})-(\ref{11-20}).

We have proven that Eqs (\ref{11-6}), (\ref{11-12}) and
(\ref{11-13}) hold for solutions with a metric $g_{\mu\nu}$ of the
form of Eqs (\ref{2-18}), (\ref{2-24}) and (\ref{2-25}) and
$J=Ma\neq0$. By the same method we can show that they also hold
for solutions with a metric of the same form and $J=0$.  The
solution given by Eqs (\ref{2-2}) and (\ref{2-3}), which is the
same with the solution A7 of the Appendix and which has $J=0$, is
obtained from our solution if we put $a=0$, make the coordinate
transformation of Eq. (\ref{8-19}) and drop the primes. Since the
coordinate transformation gives the same metric with the charges
$Q_E$ and $Q_M$ interchanged the expressions for $M$ and $dM$ of
this solution are obtained from Eqs (\ref{11-6}), (\ref{11-12})
and (\ref{11-13}) if we put $J=0$ and interchange the charges
$Q_E$ and $Q_M$ and their potentials $\phi_E$ and $\phi_M$, that
is from the relations
\begin{equation}
M=\{\frac{\pi}{N}(\frac{N}{4\pi} +2Q_E^2)(\frac{N}{4\pi} +2Q_M^2
)\}^{\frac{1}{2}} \label{11-24}
\end{equation}
\begin{equation}
M=Q_E \phi_E +Q_M \phi_M+\frac{\kappa}{4\pi}N \label{11-25}
\end{equation}
\begin{equation}
dM=\phi_E dQ_E+\phi_M dQ_M+\frac{\kappa}{8\pi}dN \label{11-26}
\end{equation}
In the above expressions the quantity $b$, the area of the outer
horizon $N$ and the surface gravity $\kappa$ are given by the
relations
\begin{equation}
b=\frac{Q_M^2-Q_E^2}{M}, \>\>\> N=4\pi r_+^H(r_+^H +b),\>\>\>
k=\frac{r_{+}^H-r_{-}^H}{2r_+^H(r_+^H+b)} \label{11-27}
\end{equation}
and the electric potential $\phi_E$ and the magnetic potential
$\phi_M$ by the relations
\begin{equation}
\phi_E=\frac{Q_E}{r_+^H}, \>\>\> \phi_M=\frac{Q_M}{r_+^H+b}
\label{11-28}
\end{equation}
which are obtained from Eq. (\ref{11-14}) if we put $a=0$ and make
the interchanges $\phi_E \rightarrow \phi_M $ and $\phi_M
\rightarrow \phi_E$.

If $a=0$ we get from Eqs (\ref{2-18}), (\ref{2-24}) and
(\ref{2-25})
\begin{equation}
g_{rr}=\frac{\rho^2}{\Delta}=\frac{r(r+b)}{(r-r_+)(r-r_-)}
\label{11-29}
\end{equation}
where $r_+>r_-$. Therefore if $r_-=-b$ the solution has only one
horizon at $r=r_+$. This is the case of the solution of STW, the
solution of GM-GHS and the solution of Schwarzschild. In this case
for solutions which are obtained from our solution for some values
of its parameters Eqs (\ref{11-6}) and (\ref{11-12})-(\ref{11-14})
hold, while for solutions which are obtained from our ``new
solution" Eqs (\ref{11-16})-(\ref{11-20}) hold.

The solution of STW is obtained from our ``new solution" for $a=0$
and $Q_E=0$ or $Q=0$ in the notation of Eq. (\ref{11-16}). If the
STW solution has electric charge $Q_E$ and magnetic charge $Q_M$
 Eqs (\ref{11-16})-(\ref{11-20}) hold with $Q=0$ and
${Q'}^2=Q_E^2+Q_M^2$. These equations give
\begin{equation}
M=\frac{1}{2}\{\frac{N}{4\pi}+2(Q_E^2+Q_M^2)\}^{\frac{1}{2}}
\label{11-30}
\end{equation}
\begin{equation}
M=Q_E \phi_E +Q_M \phi_M+\frac{\kappa}{4\pi}N \label{11-31}
\end{equation}
\begin{equation}
dM=\phi_E dQ_E+\phi_M dQ_M+\frac{\kappa}{8\pi}dN \label{11-32}
\end{equation}
\begin{equation}
 \phi_E=\frac{Q_E}{r_+}, \>\>\> \phi_M=\frac{Q_M}{r_+}, \>\>\>
b=-\frac{Q_E^2+Q_M^2}{M}\label{11-33}
\end{equation}
and since $r_-=-b=\frac{Q_E^2+Q_M^2}{M}$ Eqs (\ref{7-18}) and
(\ref{7-20}) give for the area of the horizon $N$ and the surface
gravity $\kappa$
\begin{equation}
N=4\pi r_+(r_+-\frac{Q_E^2+Q_M^2}{M}) \>\>\>\> \mbox{and} \>\>\>\>
\kappa=\frac{1}{2r_+} \label{11-34}
\end{equation}
 Eq. (\ref{11-32}) is the differential mass formula and Eq.
(\ref{11-31}) is the Smarr's type mass formula of the STW
solution.

The solution of GM-GHS is obtained from our solution for
$a=Q_E=0$. Therefore from  Eqs (\ref{11-6}) and
(\ref{11-12})-(\ref{11-13}) we get
\begin{equation}
M=\frac{1}{2}\{\frac{N}{4\pi}+2Q_M^2\}^{\frac{1}{2}} \label{11-35}
\end{equation}
\begin{equation}
M=Q_M \phi_M+\frac{\kappa}{4\pi}N \label{11-36}
\end{equation}
\begin{equation}
dM=\phi_M dQ_M+\frac{\kappa}{8\pi}dN \label{11-37}
\end{equation}
where  $r_-=-b=\frac{Q_M^2}{M}$ and
\begin{equation}
N=4\pi r_+(r_+-\frac{Q_M^2}{M}), \>\>\>\> \phi_M=\frac{Q_M}{r_+},
\>\>\>\> \kappa=\frac{1}{2r_+}\label{11-38}
\end{equation}

The solution of Schwarzschild is obtained from our solution for
$a=Q_E=Q_M=0$. Therefore since $r_- = 0$ for this solution Eqs
(\ref{11-12})-(\ref{11-13}) give the well known expressions

\begin{equation}
M=\frac{\kappa}{4\pi} N, \>\>\>\>\>\> dM=\frac{\kappa}{8\pi}dN
\label{11-39}
\end{equation}
where $N=4\pi r_+^2$, and $\kappa=\frac{1}{2r_+}$.

Consider now the case in which the metric $g_{\mu\nu}$ is of the
form of Eq. (\ref{2-18}) and $\Delta$ and $\rho^2$ are given by
Eq. (\ref{2-25'}), that is we don't have necessarily
$b=\frac{Q_E^2-Q_M^2}{M}$ and $c=2Q_M^2$. If we consider solutions
with angular momentum $J=Ma$, where $a$ is a non-zero or zero
constant and proceed as before we get instead of Eq. (\ref{11-6})
the relation
\begin{equation}
M=\{ \frac {4\pi J^2}{N}+\frac{\pi}{N}(\frac{N}{4\pi}
+c)(\frac{N}{4\pi} +c+2Mb)\}^{\frac{1}{2}} \label{11-40}
\end{equation}
If we have a solution we know $b$ and $c$ as functions of $M$ and
the charges of the solution. Therefore taking the differential of
of the above expression we get the differential mass formula of
this solution. Also if the function $M$ of Eq. (\ref{11-40}) is a
homogeneous function of its arguments we can use Euler's theorem
on homogeneous functions and get the Smarr's type mass formula of
this solution.

As an example consider the Kerr-Newman solution with arbitrary
electric charge $Q_E$ and arbitrary magnetic charge $Q_M$, which
has $b=0$ and $c=Q_E^2+Q_M^2\equiv Q^2$. Then Eq. (\ref{11-40})
for this solution takes the form
\begin{equation}
M=\{ \frac {4\pi J^2}{N}+\frac{\pi}{N}(\frac{N}{4\pi}
+Q^2)^2\}^{\frac{1}{2}} \label{11-41}
\end{equation}
and is a homogeneous function of its arguments  $x^1=J$, $x^2=N$
and $x^3=Q^2$ of degree $\frac{1}{2}$. Therefore proceeding as
before we get the following expressions for the Smarr's type mass
formula $M$ and the differential mass formula $dM$
\begin{equation}
M=2J \Omega +
\frac{Q^2r_+^H}{(r_+^H)^2+a^2}+\frac{\kappa}{4\pi}N=2J \Omega+Q_E
\phi_E +Q_M \phi_M+\frac{\kappa}{4\pi}N \label{11-42}
\end{equation}
\begin{equation}
dM=\Omega
dJ+\frac{r_+^Hd(Q^2)}{2\{(r_+^H)^2+a^2\}}+\frac{\kappa}{8\pi}dN
=\Omega dJ+\phi_E dQ_E+\phi_M dQ_M+\frac{\kappa}{8\pi}dN
\label{11-43}
\end{equation}
where the angular velocity $\Omega$ and the surface gravity
$\kappa$ are given by  Eqs (\ref{7-22}) and (\ref{7-20})
respectively and the electric potential $\phi_E$ and the magnetic
potential $\phi_M$ on the outer horizon are given by the relations
\begin{equation}
\phi_E=\frac{Q_Er_+^H}{(r_+^H)^2+a^2}, \>\>\>\> \phi_M=\frac{Q_M
r_+^H }{(r_+^H)^2+a^2} \label{11-44}
\end{equation}
We obtain the mass formulae of the solution of Kerr from Eqs
(\ref{11-41})-(\ref{11-44}) for $Q_E=Q_M=0$ and the mass formulae
of the solution of Reissner-Nordstr\"{o}m for arbitrary electric
charge $Q_E$ and magnetic charge $Q_M$ from the same equations for
$J=0$.

\appendix
\section{ Appendix\\List
of Solutions with $g_{\mu\nu}$, $F_{\mu\nu}$ and $\lambda$ of the
Form Presented in the Paper}

In this Appendix we shall present the solutions found in this
paper and a list of known solutions in order to show that these
known solutions have metric, vector field and $\lambda=\xi+i
e^{-2\phi}$ of the form of Eqs (\ref{2-18}), (\ref{3-8}) and
(\ref{9-33}) respectively and in order to make clear and simple
the comparison of our solutions with the known solutions. To make
this Appendix easier to use we have repeated in it some formulas
already existing in the paper.

All solutions of the Appendix have the metric
\begin{equation}
 g_{\mu\nu}=\left(\begin{array}{cccc}
-\frac{\Delta}{\rho^2}&0&0&\frac{a(\Delta-\rho^2)\sin^2\theta}{\rho^2}\\
0&\frac{\rho^2}{\Delta+a^2\sin^2\theta}&0&0\\
0&0&\rho^2&0\\
\frac{a(\Delta-\rho^2)\sin^2\theta}{\rho^2}&0&0&\sin^2\theta(\rho^2+2a^2\sin^2\theta)
-\frac{a^2\Delta\sin^4\theta}{\rho^2} \end{array} \right)
\label{B-1}
\end{equation}
which is a function of $a$ and two functions $\rho^2$ and
$\Delta$. The functions  $\rho^2$ and $\Delta$ are of the form
\begin{equation}
\rho^2=r(r+b)+a^2\cos^2\theta \label{B-2}
\end{equation}
\begin{equation}
\Delta=r(r+b)-2Mr+c+a^2\cos^2\theta =\rho^2-2Mr+c \label{B-3}
\end{equation}
where $M$ and $a$ are the mass and the rotation parameter of the
solution and $b$ and $c$ are constants. From the above metric we
get
\[ds^2=g_{\mu\nu}dx^{\mu}dx^{\nu}=-\frac{\Delta}{\rho^2}dt^2+
\frac{\rho^2}{\Delta+a^2\sin^2\theta}dr^2+ \rho^2d\theta^2  +
\frac{2a(\Delta-\rho^2)\sin^2\theta}{\rho^2}dtd\phi
\]
\begin{equation}
+\{\sin^2\theta(\rho^2+2a^2\sin^2\theta)
-\frac{a^2\Delta\sin^4\theta}{\rho^2}\}d\phi^2
  \label{B-4}
\end{equation}
\begin{equation}
 g^{\mu\nu}=\left(\begin{array}{cccc}
-\frac{\rho^2+2a^2\sin^2\theta
-\frac{a^2\Delta\sin^2\theta}{\rho^2}}{\Delta+a^2\sin^2\theta}&0&0
&\frac{a(\Delta-\rho^2)}{\rho^2(\Delta+a^2\sin^2\theta)}\\
0&\frac{\Delta+a^2\sin^2\theta}{\rho^2}&0&0\\
0&0&\frac{1}{\rho^2}&0\\
\frac{a(\Delta-\rho^2)}{\rho^2(\Delta+a^2\sin^2\theta)}&0&0
&\frac{\Delta}{\rho^2(\Delta+a^2\sin^2\theta)\sin^2\theta}
\end{array}\right)
\label{B-6}
\end{equation}
\begin{equation}
|g_{\mu\nu}|=\frac{1}{|g^{\mu\nu}|}=-\rho^4\sin^2\theta
\label{B-7}
\end{equation}

All solutions of this Appendix have vector field, or vector fields
if they have two such fields, which can be expressed by the
relations
\[F_{rt}=\zeta,\>\>\> F_{r\phi}=-a\sin^2\theta\zeta, \>\>\>
F_{\theta t}=-a\sin\theta\eta, \>\>\>
F_{\theta\phi}=\{r(r+b)+a^2\}\sin\theta\eta\]
\begin{equation}
F_{r\theta}=F_{t\phi}=0  \label{B-8}
\end{equation}
in terms of the rotation parameter $a$ the constant $b$ and two
functions $\zeta$ and $ \eta$.

Also if from the axion field $\xi$ and the dilaton field $\phi$
with asymptotic values $\xi_\infty=\phi_\infty=0$ we construct the
complex quantity $\lambda=\xi+i e^{-2\phi}$ all solutions of this
Appendix with $\phi\neq0$ have a $\lambda$ of the form
\begin{equation}
\lambda=\frac{i\{r+iy+(1-d)b\}}{r+iy+db} \label{B-9}
\end{equation}
where $y$ is define by the relation
\begin{equation}
 y=a\cos\theta  \label{B-10}
\end{equation}
and $d$ is a constant real or complex. For each solution we shall
give the corresponding action $S$. The solution we are talking
about is a solution of the equations of motion coming from this
action. Also we shall indicate if for this solution the rotation
parameter $a$ is vanishing or not (non-rotating or rotating
solution) and to simplify the notation we shall leave the
parameter $b$ in $\lambda$, $\xi$, $e^{-2\phi}$,
 $\zeta,\>\>
\eta,\>\> \zeta'$ and $\eta'$.

\subsection{ Our solution ({\footnotesize{ Section 6 of the
Paper}})} Action $S$
\[S = \int d^{4}x \sqrt{-g}
\{R-2\partial_{\mu}\phi\partial^{\mu}\phi-
\frac{1}{2}e^{4\phi}\partial_{\mu}\xi\partial^{\mu}\xi-
e^{-2\phi}(F_{\mu\nu}F^{\mu\nu}+F'_{\mu\nu}{F'}^{\mu\nu})\]
\begin{equation}
 -\xi(F_{\mu\nu}{F^\ast}^{\mu\nu}+F'_{\mu\nu}{F'^\ast}^{\mu\nu})\},
 \>\>\>{F^{\ast}}^{\mu\nu}=\frac{1}{2\sqrt{-g}}\epsilon^{\mu\nu\sigma\tau}F_{\sigma
\tau} \label{B-11}
\end{equation}
Parameters
\begin{equation}
 M,\>\>\>\>   a\neq0,\>\>\>\>   b=\frac{Q_E^2-Q_M^2}{M},\>\>\>\>
c=2Q_M^2,\>\>\>\>  d=0, \>\>\>\> Q_E, \>\>\>\> Q_M \label{B-12}
\end{equation}
Functions $\rho^2$, \  $\Delta$
\[\rho^2=r(r+\frac{Q_E^2-Q_M^2}{M})+y^2\]
\begin{equation}
\Delta=r(r+\frac{Q_E^2-Q_M^2}{M}) -2Mr+2Q_M^2+y^2 \label{B-13}
\end{equation}
Fields and field factors
\begin{equation}
\lambda=\frac{i(r+iy+b)}{r+iy},\>\>\>\>\> \xi=\frac{by}{r^2+y^2},
\>\>\>\>\> e^{-2\phi}=\frac{\rho^2}{r^2+y^2} \label{B-14}
\end{equation}
\begin{equation}
\zeta=\frac{Q_E(r^2-y^2)}{\rho^4},\>\>\>\>\eta=\frac{2Q_Ery}{\rho^4},\>\>\>\>
\zeta'=-\frac{Q_My(2r+b)}{\rho^4},\>\>\>\>\eta'=\frac{Q_M\{r(r+b)-y^2\}}{\rho^4}
\label{B-15}
\end{equation}
Arbitrary parameters \\
$M,\>\> a, \>\> Q_E$, and $Q_M$ : arbitrary parameters in the
metric and in the solution
\subsection{ Our ``new solution" ({\footnotesize{ Section 9 of the
Paper }})} Action $S$
\[S = \int d^{4}x \sqrt{-g}
\{R-2\partial_{\mu}\phi\partial^{\mu}\phi-
\frac{1}{2}e^{4\phi}\partial_{\mu}\xi\partial^{\mu}\xi-
e^{-2\phi}(F_{\mu\nu}F^{\mu\nu}+F'_{\mu\nu}{F'}^{\mu\nu})\]
\begin{equation}
 -\xi(F_{\mu\nu}{F^\ast}^{\mu\nu}+F'_{\mu\nu}{F'^\ast}^{\mu\nu})\},
 \>\>\>{F^{\ast}}^{\mu\nu}=\frac{1}{2\sqrt{-g}}\epsilon^{\mu\nu\sigma\tau}F_{\sigma
\tau} \label{B-11'}
\end{equation}
Parameters
\[ M, \>\>\>   a\neq0,\>\>\>   b=\frac{Q_E^2-Q_M^2}{M}, \>\>\>
c=2Q_M^2, \>\>\>  d=\frac{i\delta}{\epsilon+i\delta},\>\>\>
\delta, \>\> \epsilon,\>\> \]
\begin{equation}
\mbox{where} \>\>\> \>\>\>\>\delta^2+\epsilon^2=1, \>\> Q_E, \>\>
Q_M  \label{B-12'}
\end{equation}
Functions $\rho^2$, \  $\Delta$
\begin{equation}
\rho^2=r(r+\frac{Q_E^2-Q_M^2}{M})+y^2,\>\>\>
\Delta=r(r+\frac{Q_E^2-Q_M^2}{M}) -2Mr+2Q_M^2+y^2 \label{B-13'}
\end{equation}
Fields and field factors
\begin{equation}
\lambda=\frac{i\{(\epsilon+i\delta)(r+iy)+\epsilon b\}}{(\epsilon
+i\delta)(r+iy)+i\delta b}\label{B-14'}
\end{equation}
\begin{equation}
\xi=\frac{b\{\delta\epsilon(2r+b)+(\epsilon^2-\delta^2)y\}}{\delta^2\{(r+b)^2+y^2\}
+\epsilon^2(r^2+y^2)+2\delta\epsilon b y}\label{B-15'}
\end{equation}
\begin{equation}
e^{-2\phi}=\frac{\rho^2}{\delta^2\{(r+b)^2+y^2\}
+\epsilon^2(r^2+y^2)+2\delta\epsilon b y}\label{B-16}
\end{equation}
\begin{equation}
\zeta=\frac{Q_E}{\rho^4}\{\delta y(2r+b)-\epsilon (r^2-y^2)\},
 \>\>\>\>\eta=-\frac{Q_E}{\rho^4}\{\delta(r(r+b)-y^2)+2\epsilon ry\}
\label{B-17}
\end{equation}
\begin{equation}
\zeta'=\frac{Q_M}{\rho^4}\{\delta((r+b)^2-y^2)+\epsilon(2r+b)y\},
\>\>\>\>\eta'=\frac{Q_M}{\rho^4}\{2\delta y(r+b)-\epsilon(r(r+b)-y^2)\}
\label{B-18}
\end{equation}
Arbitrary parameters \\
$M,\>\>\> a,\>\>\> Q_E, \>\>\>Q_M $ : arbitrary parameters of the metric\\
$M, \>\>\>a,\>\>\>  Q_E,\>\>\> Q_M,\>\>\> \delta$ : arbitrary parameters of the solution\\
The quantities $\lambda'',\>\> \xi'',\>\> \phi'',\>\> \sigma,\>\>
\tau,\>\> \sigma',\>\> \tau'$ of Section 9 were renamed
$\lambda,\>\> \xi,\>\> \phi, \zeta, \eta, \>\>\zeta'$ and $\eta'$
here.

\subsection{Solution of Sen with electric charge $Q_E$ only ({\footnotesize{Phys.
 Rev. Lett. {\bf{69}}, 1006 (1992)}})}
Action $S$
\[S = \int d^{4}x \sqrt{-g}
\{R-2\partial_{\mu}\phi\partial^{\mu}\phi-
\frac{1}{2}e^{4\phi}\partial_{\mu}\xi\partial^{\mu}\xi-
e^{-2\phi}F_{\mu\nu}F^{\mu\nu}\]
\begin{equation}
 -\xi F_{\mu\nu}{F^\ast}^{\mu\nu}\},
 \>\>\>{F^{\ast}}^{\mu\nu}=\frac{1}{2\sqrt{-g}}\epsilon^{\mu\nu\sigma\tau}F_{\sigma
\tau} \label{B-19}
\end{equation}
Parameters
\begin{equation}
 M,\>\>\>\>   a\neq0,\>\>\>\>   b=\frac{Q_E^2}{M},\>\>\>\>
c=0,\>\>\>\>  d=0, \>\>\>\> Q_E \label{B-20}
\end{equation}
Functions $\rho^2$, \  $\Delta$
\begin{equation}
\rho^2=r(r+\frac{Q_E^2}{M})+y^2,\>\>\>
\Delta=r(r+\frac{Q_E^2}{M})-2Mr+y^2 \label{B-21}
\end{equation}
Fields and field factors
\begin{equation}
\lambda=\frac{i(r+iy+b)}{r+iy},\>\>\>\>\> \xi=\frac{by}{r^2+y^2},
\>\>\>\>\> e^{-2\phi}=\frac{\rho^2}{r^2+y^2} \label{B-22}
\end{equation}
\begin{equation}
\zeta=\frac{Q_E(r^2-y^2)}{\rho^4},\>\>\>\>\eta=\frac{2Q_Ery}{\rho^4}\label{B-23}
\end{equation}
Arbitrary parameters \\
$M,\>\> a, \>\> Q_E$ : arbitrary parameters in the
metric and in the solution

\subsection{Solution of Sen with electric charge $Q_E$ and magnetic charge $Q_M$
({\footnotesize{ Ref. \cite {Se2}. The complete solution is not
given explicitly}})} Action $S$
\[S = \int d^{4}x \sqrt{-g}
\{R-2\partial_{\mu}\phi\partial^{\mu}\phi-
\frac{1}{2}e^{4\phi}\partial_{\mu}\xi\partial^{\mu}\xi-
e^{-2\phi}F_{\mu\nu}F^{\mu\nu}\]
\begin{equation}
 -\xi F_{\mu\nu}{F^\ast}^{\mu\nu}\},
 \>\>\>{F^{\ast}}^{\mu\nu}=\frac{1}{2\sqrt{-g}}\epsilon^{\mu\nu\sigma\tau}F_{\sigma
\tau} \label{B-24}
\end{equation}
Parameters
\begin{equation}
 M,\>\>\>\>   a\neq0,\>\>\>\>   b=\frac{Q_E^2+Q_M^2}{M},\>\>\>\>
c=0,\>\>\>\>  d=\frac{iQ_M}{Q_E+iQ_M}, \>\>\>\> Q_E, \>\>\>\> Q_M
\label{B-25}
\end{equation}
Functions $\rho^2$, \  $\Delta$
\begin{equation}
\rho^2=r(r+\frac{Q_E^2+Q_M^2}{M})+y^2,\>\>\>
\Delta=r(r+\frac{Q_E^2+Q_M^2}{M})-2Mr+y^2 \label{B-26}
\end{equation}
Fields and field factors
\begin{equation}
\lambda=\frac{i\{(Q_E+iQ_M)(r+iy)+Q_E b\}}{(Q_E
+iQ_M)(r+iy)+iQ_M b}\label{B-27}
\end{equation}
\begin{equation}
\xi=\frac{b\{Q_E Q_M(2r+b)+(Q_E^2-Q_M^2)y\}}{Q_M^2\{(r+b)^2+y^2\}
+Q_E^2(r^2+y^2)+2Q_E Q_M b y}\label{B-28}
\end{equation}
\begin{equation}
e^{-2\phi}=\frac{(Q_E^2+Q_M^2)\rho^2}{Q_M^2\{(r+b)^2+y^2\}
+Q_E^2(r^2+y^2)+2Q_E Q_M b y}\label{B-29}
\end{equation}
\begin{equation}
\zeta=\frac{Q_E(r^2-y^2)-Q_M(2r+b)y}{\rho^4},
 \>\>\>\>\eta=\frac{Q_M\{r(r+b)-y^2\}+2Q_E r y}{\rho^4}
\label{B-30}
\end{equation}
Arbitrary parameters \\
$M,\>\> a, \>\> Q_E^2+Q_M^2$ : arbitrary parameters of the metric\\
$M,\>\> a, \>\> Q_E,\>\>Q_M $ : arbitrary parameters of the solution

\subsection{ Kerr-Newman solution for electric charge $Q_E$
and magnetic charge $Q_M$ ({\footnotesize{ Ref. \cite{Ca1}}})}
Action $S$
\begin{equation}
S = \int d^{4}x \sqrt{-g}(R-F_{\mu\nu}F^{\mu\nu})\label{B-31}
\end{equation}
Parameters
\begin{equation}
 M,\>\>\>\>   a\neq0,\>\>\>\>   b=0,\>\>\>\>
c=Q_E^2+Q_M^2,\>\>\> Q_E,\>\>\> Q_M\label{B-32}
\end{equation}
Functions $\rho^2$, \  $\Delta$
\begin{equation}
\rho^2=r^2+y^2, \>\>\> \Delta=r^2 -2Mr+Q_E^2+Q_M^2+y^2
\label{B-33}
\end{equation}
Field factors
\begin{equation}
\zeta=\frac{Q_E(r^2-y^2)-2Q_Mry}{\rho^4},
 \>\>\>\>\eta=\frac{Q_M(r^2-y^2)+2Q_E r y}{\rho^4}
\label{B-34}
\end{equation}
Arbitrary parameters\\
$ M, \>\>\> a, \>\>\> Q_E^2+Q_M^2 $ : arbitrary parameters of the metric\\
$ M, \>\>\> a, \>\>\> Q_E, \>\>\> Q_M $ : arbitrary parameters of the solution

\subsection{Solution of Kerr ({\footnotesize{Phys. Rev. Lett. {\bf{11}}, 237 (1963)}})}
Action $S$
\begin{equation}
S = \int d^{4}x \sqrt{-g}R \label{B-35}
\end{equation}
Parameters
\begin{equation}
 M,\>\>\>\>   a\neq0,\>\>\>\>   b=0,\>\>\>\>
c=0\label{B-36}
\end{equation}
Functions $\rho^2$, \  $\Delta$
\begin{equation}
\rho^2=r^2+y^2, \>\>\> \Delta=r^2 -2Mr+y^2 \label{B-37}
\end{equation}
Arbitrary parameters\\
$ M, \>\>\> a, \>\>\> $ : arbitrary parameters in the metric and in the solution

\subsection{Solution given by Eqs  (\ref{2-2}) and (\ref{2-3}) of this
paper\\
 ({\footnotesize{ Class. Quant. Grav. {\bf{23}}, 7591 (2006) Eqs (54)-(57)
  for $\psi_{0}=0$, $AB=2Q_E^2$, $(\alpha-A)(\alpha-B)=2Q_M^2$, $2M=\alpha-A-B$,
$\psi=2\phi$ and $\alpha=b$
   }})}
Action $S$
\begin{equation}
S = \int d^{4}x \sqrt{-g}
\{R-2\partial_{\mu}\phi\partial^{\mu}\phi
-e^{-2\phi}F_{\mu\nu}F^{\mu\nu}\}
 \label{B-38}
\end{equation}
Parameters
\begin{equation}
 M,\>\>\>\>   a=0,\>\>\>\>   b=\frac{Q_M^2-Q_E^2}{M},\>\>\>\>
c=2Q_E^2,\>\>\>\>  d=1, \>\>\>\> Q_E, \>\>\>\> Q_M \label{B-39}
\end{equation}
Functions $\rho^2$, \  $\Delta$
\begin{equation}
\rho^2=r(r+\frac{Q_M^2-Q_E^2}{M}),\>\>\>
\Delta=r(r+\frac{Q_M^2-Q_E^2}{M})-2Mr+2Q_E^2
 \label{B-40}
\end{equation}
Fields and field factors
\begin{equation}
\xi=0,\>\>\>\>\>\>\>\>  e^{2\phi}=1+\frac{b}{r}\label{B-40'}
\end{equation}
\begin{equation}
\zeta=\frac{Q_E}{r^2},\>\>\>\>\>\>\>\>\eta=\frac{Q_M}{\rho^2}\label{B-41}
\end{equation}
Arbitrary parameters \\
$M,\>\> Q_E, \>\> Q_M$ : arbitrary parameters in  the
metric and in the solution\\
Line element
\[ds^2=-\{1-\frac{2M}{r}+\frac{2Q_M^2M}{r(Mr+Q_M^2-Q_E^2)}\}dt^2
+\{1-\frac{2M}{r}+\frac{2Q_M^2M}{r(Mr+Q_M^2-Q_E^2)}\}^{-1}dr^2\]
\begin{equation}
+r(r +\frac{Q_M^2-Q_E^2}{M})(d\theta^2+\sin^2\theta d\phi^2)
\label{B-41'}
\end{equation}

The solution takes simpler form in terms of the parameters $A, \>
B, \>  b$ where $ 2Q_E^2=AB, \>\>\> 2Q_M^2=(b-A)(b-B), \>\>\>
2M=b-A-B.$\\ Functions $\rho^2$, $\Delta$
\begin{equation}
\rho^2= r(r +b), \>\>\> \Delta=(r+A)(r+B) \label{B-41''}
\end{equation}
Fields and field factors
\begin{equation}
\xi=0,\>\>\>\>\>\>\>\>  e^{2\phi}=1+\frac{b}{r}\label{B-41'''}
\end{equation}
\begin{equation}
\zeta=\frac{\sqrt{AB}}{\sqrt{2}r^2},\>\>\>\>\>\>\>\>
\eta=\frac{\sqrt{(b-A)(b-B)}}{\sqrt{2}\rho^2}\label{B-42'}
\end{equation}
Arbitrary parameters \\
$A<0,\>\> B<0, \>\> b>0$ : arbitrary parameters in  the
metric and in the solution\\
Line element
\begin{equation}
ds^{2}=-\frac{(r+A)(r+B)}{r(r+b)}dt^{2}
+\frac{r(r+b)}{(r+A)(r+B)}dr^{2} +r(r+b)(d\theta^{2}+sin^{2}\theta
{d{\phi}^{2}})
 \label{B-42''}
\end{equation}

\subsection{Solution of Shapere, Trivedi and Wilczek ({\footnotesize{Mod.
 Phys. Lett. {\bf{A6}},2677 (1991)}})}
Action $S$
\[S = \int d^{4}x \sqrt{-g}
\{R-2\partial_{\mu}\phi\partial^{\mu}\phi-
\frac{1}{2}e^{4\phi}\partial_{\mu}\xi\partial^{\mu}\xi-
e^{-2\phi}F_{\mu\nu}F^{\mu\nu}\]
\begin{equation}
 -\xi F_{\mu\nu}{F^\ast}^{\mu\nu}\},
 \>\>\>{F^{\ast}}^{\mu\nu}=\frac{1}{2\sqrt{-g}}\epsilon^{\mu\nu\sigma\tau}F_{\sigma
\tau} \label{B-42}
\end{equation}
Parameters
\begin{equation}
 M,\>\>\>\>   a=0,\>\>\>\>   b=-\frac{Q_E^2+Q_M^2}{M},\>\>\>\>
c=2(Q_E^2+Q_M^2),\>\>\>\>  d=-\frac{iQ_E}{Q_M-iQ_E}, \>\>\>\> Q_E,
\>\>\>\> Q_M \label{B-43}
\end{equation}
Functions $\rho^2$, \  $\Delta$
\begin{equation}
\rho^2=r(r-\frac{Q_E^2+Q_M^2}{M}),\>\>\> \label{B-44}
\end{equation}
\begin{equation}
\Delta=r(r-\frac{Q_E^2+Q_M^2
}{M})-2Mr+2(Q_E^2+Q_M^2)=(r-2M)(r-\frac{Q_E^2+Q_M^2 }{M})
\label{B-44'}
\end{equation}
Fields and field factors \cite{Sh1}
\begin{equation}
\lambda=\frac{i\{(Q_M-iQ_E)r+Q_M b)\}}{(Q_M-iQ_E)r-iQ_E
b}\label{B-45}
\end{equation}
\begin{equation}
\xi=-\frac{Q_E Q_M b(2r+b)}{Q_E^2(r+b)^2+Q_M^2 r^2}, \>\>\>\>
e^{-2\phi}=\frac{(Q_E^2+Q_M^2)\rho^2}{Q_E^2(r+b)^2+Q_M^2 r^2}
\label{B-45'}
\end{equation}
\begin{equation}
\zeta=\frac{Q_E}{r^2},\>\>\>\>\eta=\frac{Q_M}{\rho^2}\label{B-46}
\end{equation}
Arbitrary parameters \\
$M,\>\>\> Q_E^2+Q_M^2$ : arbitrary parameters of the metric\\
$M, \>\>\> Q_E, \>\>\> Q_M $  :  arbitrary parameters of the solution\\
Line element
\begin{equation}
dt^2=-(1-\frac{2M}{r})dt^2+(1-\frac{2M}{r})^{-1}dr^2+r(r
-\frac{Q_E^2+Q_M^2}{M})(d\theta^2+\sin^2\theta d\phi^2)
\label{B-47}
\end{equation}

\subsection{ GM-GHS Solution
 ({\footnotesize{G. Gibbons, Nucl. Phys. {\bf{B207}}, 337 (1982); G. Gibbons and
Maeda  Nucl. Phys. {\bf{B298}}, 741 (1988); D. Garfinkle, G.
Horowitz and A. Strominger, Phys. Rev. {\bf{D43}}, 3140 (1991) for
$\phi_0=0$; Erratum Phys. Rev. {\bf{D45}}, 3888 (1992) }})}
 Action $S$
\begin{equation}
S = \int d^{4}x \sqrt{-g}
\{R-2\partial_{\mu}\phi\partial^{\mu}\phi
-e^{-2\phi}F_{\mu\nu}F^{\mu\nu}\}
 \label{B-48}
\end{equation}
Parameters
\begin{equation}
 M,\>\>\>\>   a=0,\>\>\>\>   b=-\frac{Q_M^2}{M},\>\>\>\>
c=2Q_M^2,\>\>\>\>  d=0, \>\>\>\> Q_M \label{B-49}
\end{equation}
Functions $\rho^2$, $\Delta$

\[\rho^2=r(r-\frac{Q_M^2}{M})\]
\begin{equation}
\Delta=r(r-\frac{Q_M^2}{M})-2Mr+2Q_M^2
=(r-2M)(r-\frac{Q_M^2}{M})\label{B-50}
\end{equation}
Fields and field factors
\begin{equation}
\xi=0,\>\>\>\>\>  e^{-2\phi}=1+\frac{b}{r}, \>\>\>\>
\zeta=0,\>\>\>\>\eta=\frac{Q_M}{\rho^2}\label{B-51}
\end{equation}
Arbitrary parameters \\
$M,\>\>\>\>\>\> Q_M$ : arbitrary parameters in  the
metric and in the solution\\
Line element
\begin{equation}
dt^2=-(1-\frac{2M}{r})dt^2+(1-\frac{2M}{r})^{-1}dr^2+r(r
-\frac{Q_M^2}{M})(d\theta^2+\sin^2\theta d\phi^2)
\label{B-52}
\end{equation}

\subsection{ Reissner-Nordstr\"{o}m  solution for electric charge $Q_E$
and magnetic charge $Q_M$ } Action $S$
\begin{equation}
S = \int d^{4}x \sqrt{-g}(R-F_{\mu\nu}F^{\mu\nu})\label{B-53}
\end{equation}
Parameters
\begin{equation}
 M,\>\>\>\>   a=0,\>\>\>\>   b=0,\>\>\>\>
c=Q_E^2+Q_M^2,\>\>\> Q_E,\>\>\> Q_M\label{B-54}
\end{equation}
Functions $\rho^2$, \  $\Delta$
\begin{equation}
\rho^2=r^2, \>\>\>
\Delta=r^2 -2Mr+Q_E^2+Q_M^2
\label{B-55}
\end{equation}
Field factors
\begin{equation}
\zeta=\frac{Q_E}{r^2},
 \>\>\>\>\eta=\frac{Q_M}{r^2}
\label{B-56}
\end{equation}
Arbitrary parameters\\
$ M, \>\>\>\>\> Q_E^2+Q_M^2 $ : arbitrary parameters of the metric\\
$ M, \>\>\>\>\> Q_E, \>\>\>\> Q_M $ : arbitrary parameters of the solution\\
Line element
\begin{equation}
ds^2=-(1-\frac{2M}{r}+\frac{Q_E^2+Q_M^2}{r^2})dt^2+(1-\frac{2M}{r}+\frac{Q_E^2
+Q_M^2}{r^2})^{-1}dr^2+r^2(d\theta^2+\sin^2\theta d\phi^2)
\label{B-57}
\end{equation}

\subsection{Schwarzschild  solution}
Action $S$
\begin{equation}
S = \int d^{4}x \sqrt{-g}R\label{B-58}
\end{equation}
Parameters
\begin{equation}
 M,\>\>\>\>   a=0,\>\>\>\>   b=0,\>\>\>\>
c=0\label{B-59}
\end{equation}
Functions $\rho^2$, \  $\Delta$
\begin{equation}
\rho^2=r^2, \>\>\>\>\>\>
\Delta=r^2 -2Mr
\label{B-60}
\end{equation}
Line element
\begin{equation}
dt^2=-(1-\frac{2M}{r})dt^2+(1-\frac{2M}{r})^{-1}dr^2+r^2(d\theta^2
+\sin^2\theta d\phi^2)
\label{B-61}
\end{equation}


\begin{thebibliography}{99}
\bibitem{Ke1} R. P. Kerr, Phys. Rev. Lett. {\bf{11}}, 237 (1963)
\bibitem{Ne1} A. J. Janis and E. T. Newman, J. Math. Phys.  {\bf{6}}, 915 (1965)
\bibitem{Ne2} E. T. Newman, E. Couch, K. Chinnapared, A. Exton, A. Prakash and R.
Torrence, J. Math. Phys.  {\bf{6}}, 918 (1965)
\bibitem{Se1} Ashore Sen, Phys. Rev. Lett. {\bf{69}}, 1006 (1992)
\bibitem{Gi1} G. Gibbons, Nucl. Phys. {\bf{B207}}, 337 (1982); G. Gibbons and
Maeda Nucl. Phys. {\bf{B298}}, 741 (1988); D. Garfinkle, G.
Horowitz and A. Strominger, Phys. Rev. {\bf{D43}}, 3140 (1991);
Erratum Phys. Rev. {\bf{D45}}, 3888 (1992)
\bibitem{Ya1} S. Yazadjiev, Gen. Rel. Grav. {\bf{32}}, 2345 (2000)
\bibitem{Ky1} E. Kyriakopoulos, Int. J. Mod. Phys. {\bf{D 15}}, 2223 (2006)
\bibitem{Ky2} E. Kyriakopoulos, gr-gc/ 0611060
\bibitem{Ky3} E. Kyriakopoulos, Class. Quant. Grav. {\bf{23}}, 7591 (2006)
\bibitem{Be1} E. Bergshoeff, R. Kallosh and T. Ort\'{i}n,
Nucl. Phys. {\bf{B478}}, 156 (1996)
\bibitem{Ra1} Ray d'Inverno, {\em Introducing Einstein's
Relativity } ( Clarendon Press, Oxford 1995)
\bibitem{Ca1} B. Carter in {\em  Black Holes}, Les Houches
Ao\^{u}t 1972, Edited by C. DeWitt and B. S. DeWitt, (Gordon and
Breach Science Publishers, New York 1973)
\bibitem{Se2} Ashoke Sen,  Nucl. Phys. {\bf{B404}}, 109 (1993);
Talks given at the John Hopkins Workshop held at G\"{o}tenborg
June 8-10, 1992 and ICTP Summer Workshop held at Trieste, July 2-3
1992, hep-th/9210050. The external field strength hair present for
black holes with angular momentum and electric and magnetic charge
in string gravity were computed to first order in the string
tension by B. Campbell, N. Kaloper and K. Olive, Phys. Lett.
{\bf{B285}}, 199 (1991)
\bibitem{Sh1} A. Shapere, S. Trivedi and F. Wilczek
Mod. Phys. Lett. {\bf{A6}}, 2677 (1991) The axion field of this
solution and the axion field of Eq. (\ref{10-14}) have opposite
signs. This sign difference is corrected if we take in our paper
$\epsilon^{0123}=-1$. The axion field of the solution A8 of the
Appendix has the sign of the axion field of Eq. (\ref{10-14})
\bibitem{Sm1} L. Smarr Phys. Rev. Lett. {\bf{30}}, 71 (1973)
\bibitem{To1} To make the notation simpler the function
$(\Delta)_{paper}$ of the present paper is slightly different from
the function $(\Delta)_{MTW}$ of the book of C. W. Misner, K. S.
Thorne and J. A. Wheeler, {\em 1973 Gravitation}, (W. H. Freeman
and Company, San Francisco ) and of several other books, which
define $\Delta$ as in the book of C. W. Misner et. al. It is $
(\Delta)_{MTW}=(\Delta)_{paper} +a^2\sin^2\theta $.
\bibitem{Bo1} The computation of the Ricci tensor $R_{\mu\nu}$, the Ricci
scalar $R$ and the curvature scalar
$R_{\mu\nu\sigma\tau}R^{\mu\nu\sigma\tau}$ was done with the help
of a program given to me by S. Bonanos, whom I thank.
\bibitem{Ho1} James H. Horne and Gary T. Horowitz, Phys. Rev. {\bf{D46}}, 1340 (1992)
\bibitem{Ad1} R. Adler, M. Bazin and M. Schiffer, {\em
Introduction to General Relativity, Second Edition},( McGraw-Hill
Book Company, New York 1975)
\bibitem{Ga1} D. V. Gal'tsov and P. S. Letelier, Phys. Rev. {\bf{D55}}, 3580 (1997)
\bibitem{Ha1} S. W. Hawking and G. F. R. Ellis {\em The Large Scale Structure
of Space-Time} (Cambridge University Press, Cambridge 1973 )

\end{thebibliography}
\end{document}